\documentclass[ApJS,twocolumn]{aastex631}
\usepackage{amssymb, amsmath, graphicx, dcolumn, color,units, xspace, mathtools, physics, tensor, bm, lipsum, revsymb, float}
\usepackage[normalem]{ulem}

\makeatletter
\let\frontmatter@title@above=\relax
\makeatother

\usepackage{hyperref}
\usepackage{booktabs,dcolumn}

\begin{document}

\title{Significant challenges for astrophysical inference with next-generation gravitational-wave observatories}

\author{A. Makai Baker}
\affiliation{School of Physics and Astronomy, Monash University, VIC 3800, Australia}
\affiliation{OzGrav: The ARC Centre of Excellence for Gravitational-Wave Discovery, Clayton, VIC 3800, Australia}

\author{Paul D. Lasky}
\affiliation{School of Physics and Astronomy, Monash University, VIC 3800, Australia}
\affiliation{OzGrav: The ARC Centre of Excellence for Gravitational-Wave Discovery, Clayton, VIC 3800, Australia}

\author{Eric Thrane}
\affiliation{School of Physics and Astronomy, Monash University, VIC 3800, Australia}
\affiliation{OzGrav: The ARC Centre of Excellence for Gravitational-Wave Discovery, Clayton, VIC 3800, Australia}

\author{Jacob Golomb}
\affiliation{Department of Physics, California Institute of Technology, Pasadena, California 91125, USA}
\affiliation{LIGO Laboratory, California Institute of Technology, Pasadena, California 91125, USA}

\begin{abstract}
    The next generation of gravitational-wave observatories will achieve unprecedented strain sensitivities with an expanded observing band.
    They will detect ${\cal O}(10^5)$ binary neutron star (BNS) mergers every year, the loudest of which will be in the band for $\approx 90$ minutes with signal-to-noise ratios $\approx 1500$.
    Current techniques will not be able to determine the astrophysical parameters of the loudest of next-gen BNS signals.
    We show that subtleties arising from the rotation of the Earth and the free-spectral range of gravitational-wave interferometers dramatically increases the complexity of next-gen BNS signals compared to the one-minute signals seen by LIGO--Virgo.
    Various compression methods currently relied upon to speed up the most expensive BNS calculations---reduced-order quadrature, multi-banding, and relative binning---will no longer be effective. 
    We carry out reduced-order inference on a simulated next-gen BNS signal taking into account the Earth's rotation and the observatories' free-spectral range.
    We show that standard data compression techniques become impractical, and the full problem becomes computationally infeasible, when we include data below $\approx\unit[16]{Hz}$---a part of the observing band that is critical for precise sky localisation. 
    We discuss potential paths towards solving this complex problem. 
\end{abstract}

\section{Introduction}
Preparations are underway for the next generation of terrestrial gravitational-wave observatories.
The European-based Einstein Telescope \citep{Punturo_2010} is designed to achieve strain sensitivities of $\gtrsim\unit[2\times10^{-25}]{Hz^{-1/2}}$ with a minimum observing frequency of $\unit[1]{Hz}$ while the American-led Cosmic Explorer \citep{CE} is designed to achieve a similar strain sensitivity to the Einstein Telescope with a minimum frequency of $\unit[5]{Hz}$ \citep{Hall_2021}.
These next-gen facilities will revolutionise gravitational-wave astronomy.
Whereas LIGO--Virgo have so far announced ${\cal O}(100)$ gravitational-wave observations \citep{gwtc1, gwtc2, gwtc3}, Cosmic Explorer and the Einstein Telescope will detect most stellar mass binary black hole merger in the Universe and all but the most distant binary neutron star mergers---about $10^5$ every year \citep{CE}.

Expanding the observing band below the $\approx\unit[20]{Hz}$ boundary of current detectors will unlock new science.
For example, next-gen observatories will become sensitive to intermediate mass black holes \citep{Greene}, and will observe binary neutron stars for $\approx\unit[90]{minutes}$ compared to the $\lesssim\unit[1]{minute}$ signals observed in LIGO--Virgo.
This increased duration can be leveraged to improve the localisation of binary neutron stars.
During a $\unit[90]{minute}$ signal, the detectors move a distance of $\approx´\unit[1700]{km}$ due to the Earth's rotation. 
This creates a baseline comparable to the $\approx\unit[3000]{km}$ distance between LIGO Hanford and LIGO Livingston, which can be used for triangulation with a single detector.
Taking advantage of this baseline, a single Cosmic Explorer can localise a binary neutron star to within $\approx \unit[10]{deg^2}$ \citep{baral}, which is in contrast to the localisation of $\gtrsim \unit[8000]{deg^2}$ that has been achieved for events like GW190425, seen by a single LIGO detector \citep{gw190425}.\footnote{There is no substitute for networks of \textit{two} or more next-gen observatories, which can localise a large fraction of their binary neutron star mergers to $\lesssim \unit[1]{deg^2}$; see, e.g., \cite{Gardner}.}
Next-gen observatories will also dramatically increase the signal-to-noise ratio (SNR) for binary neutron star signals from SNR=40 for GW170817 \citep{GW170817} to $\text{SNR}\approx1600$ for a GW170817-like event as seen by Cosmic Explorer.

In order to measure the parameters of binary neutron stars---such as their sky location, their masses, etc---gravitational-wave astronomers rely on Bayesian inference; see, e.g., \cite{thrane_talbot_2019}.
Bayesian inference is already a serious computational expense for LIGO--Virgo; each event can take between a hours to days to analyse, depending on various factors such as the duration of the signal, the SNR, and the computational cost of the waveform used in the analysis.
Relatively longer waveforms require more computations than shorter waveforms, causing inference to take longer.
Somewhat counterintuitively, analyzing signals with higher SNRs takes longer because stochastic samplers take longer to converge on the narrow likelihood peak.
\cite{Hu} recently pointed out that the large number of long-duration, high-SNR events observed by next-gen detectors likely require a computing paradigm shift.

Data compression techniques---such as reduced-order quadrature \citep{Canizares_2013}, multi-banding \citep{multibanding}, and relative binning \citep{Cornish:2010kf,modebymode_relativebinning}---have emerged as essential tools in the battle to control computational costs.
The basic idea with all of these methods is to exploit sparsity in the gravitational-wave time series. Multi-banding and relative binning use  different techniques to coarsely sample the time series, whereas reduced-order quadrature represents the series in terms of a comparatively small number of basis elements.
This can sometimes dramatically reduce the number of computations required to estimate the gravitational-wave likelihood function, leading to remarkable speed-ups.\footnote{Another way to understand these compression methods is as follows. The space of time series that correspond to gravitational-wave signals is only a small subset of the space of all time series. Therefore, one can sometimes compress the information in a gravitational-wave time series into a much smaller format.}
For example, LIGO--Virgo regularly use reduced order models to speed up binary neutron star calculations by a factor of $\approx$150, allowing researchers to obtain results in the span of a few days what could otherwise take a year \citep{Canizares_2013}.

In \cite{Smith_2016}, some of us argued that reduced order modelling could be used to manage the immense cost of analysing a $\unit[90]{minute}$-long binary neutron star event like GW170817 \citep{GW170817} observed by a network of next-gen observatories with $\text{SNR}>1000$.
That demonstration was not cheap, requiring ten hours using $160$ high-interconnect cores that were modern in 2021.
However, the analysis would have been hopeless without the $10^4$ speed-up made possible with reduced order methods.
Unfortunately, we now believe that demonstration was missing key features.

First, we did not properly take into account  effects that arise from the rotation of the Earth over the duration of the signal and the lower free spectral range of next-gen detectors.
The free spectral range is the frequency corresponding to the inverse travel time of light in the interferometer arms.
The response of the interferometer to gravitational waves with frequencies near the free spectral range is complicated: the amplitude and phase are modulated depending on the direction of the incident waves \citep{Rakhmanov:2008is,Essick_2017}.
The effect is largely ignored for LIGO--Virgo, which have a free spectral range frequency of $f_{\rm fsr} \approx\unit[38]{kHz}$---safely above the spectral content of astrophysical gravitational waves.
However, the long arms of next-gen observatories such as the 40\,km Cosmic Explorer place the free spectral range at $\approx\unit[3750]{Hz}$, which is low enough (it turns out) to significantly affect the response of the detectors to binary neutron star signals.
Given that the effect of the free spectral range depends on the direction of the incident gravitational wave, the combination of the free spectral range and rotation of the Earth create a more complicated signal space than we considered in \cite{Smith_2016}.

Second, work by \cite{Morisaki_2023} has provided more stringent guidelines for determining if a reduced order model is sufficiently accurate to be used for unbiased inference calculations. 
Previous work had used the peak of the mismatch between the waveform and reduced-order model evaluated over many parameters to determine whether the reduced-order quadrature likelihood was accurate enough \cite{rory}. However, \cite{Morisaki_2023} suggests that the accuracy of the reduced-order quadrature rule is determined by the relative error in the log likelihood ratio.\footnote{To make this relation explicit, we derive an approximate relation between the relative error in the log likelihood ratio and the maximum signal-to-noise ratio that can be studied with reduced-order quadrature without introducing bias in the posteriors. 
The relation is given in Sec.~\ref{sec:accuracy_of_roq} and derived in Appendix~\ref{appendix:aprior}.}

In this paper, we seek to test the use of compression methods like reduced order modelling on next-gen gravitational-wave detectors---this time carefully taking into account the rotation of the Earth, the free spectral range, and the strict accuracy requirements from \cite{Morisaki_2023}.
We show that---in practice---reduced order techniques will not work for binary neutron star signals in next-gen observatories because the size of the reduced order model becomes unwieldy.
We argue that data compression techniques are in general unlikely to provide a solution.
We discuss various alternate pathways worthy of exploration.

The remainder of this paper is organised as follows.
In Section~\ref{sec:building_roms} we review the formalism of reduced order methods.
In doing so, we aim to demystify for a broad audience what is sometimes regarded as an esoteric topic.
In Section~\ref{sec:alternatives}, we discuss other forms of compression and argue that they are inferior to reduced order methods, which justifies our focus on reduced order models.
In Section~\ref{sec:methodology}, we describe our method for determining the computational requirements for binary neutron stars observed by next-gen observatories---taking into account the free spectral range and the rotation of the Earth.
(Impatient readers can skip directly to Fig.~\ref{fig:snr_curve}, which shows the GW170817-like-event signal-to-noise ratio achievable with reduced order models---given a fixed memory allotment of $\unit[100]{Gb}$.)
In Section~\ref{sec:inference}, we demonstrate Bayesian inference with one of our reduced order models.
(We employ a minimum frequency of $\unit[26]{Hz}$ in order to control the size of the model.)
Finally in Section~\ref{sec:conclusions}, we summarise our results and discuss possible solutions.

This paper also includes extensive appendices.
In Appendix~\ref{appendix:rombus}, we describe how to use the open access software \texttt{rombus},\footnote{\url{https://github.com/ADACS-Australia/rombus}} designed to make it easy for non-specialists to generate reduced order models.
In Appendix~\ref{sec:jacob}, we describe our implementation of time- and frequency-dependent effects needed for analysis of signals in next-gen detectors.\footnote{\url{https://git.ligo.org/lscsoft/bilby/-/merge_requests/1370}}

\section{Formalism}\label{sec:building_roms}
In this Section, we describe step-by-step the construction of a reduced order model for a binary neutron star waveform. 
We then introduce the reduced order quadrature likelihood and describe techniques for efficient reduced order model construction. 
We finish with a discussion of the Earth's rotation and the free spectral range and describe why these increase the size of our reduced model.
This section is included to provide an introduction for non-experts.
Readers interested in the punchline should consider skipping ahead to Section~\ref{sec:methodology}.
The figures referenced in this section are included in Appendix~\ref{app:plots} so as to keep the main body relatively succinct.
The \texttt{rombus} python package is used to construct the reduced order models described in this work. 
To build these models, we make a number of improvements to \texttt{rombus} as described in Appendix~\ref{appendix:rombus}.

\subsection{Building Reduced Order Models}\label{sec:building_rom}
Our goal is to efficiently calculate a gravitational-wave likelihood that compares strain data $d$ with a template for the observed strain $h$
\begin{align}\label{eq:template}
    h(t) = \sum_A F_A\big(t + \Delta(t)\big) h_A\big(t+\Delta(t)\big) .
\end{align}
Here, $A=+,\times$ is an index for the gravitational-wave polarisation state, $h_A$ is the gravitational wave for polarisation state $A$ and $F_A$ is an antenna response factor.
The antenna response factors implicitly depend on the location $\hat\Omega$ and polarisation $\psi$ of the source.
For long signals, the antenna response factors vary over the duration of the signal because the detector-frame direction to the source changes as the Earth rotates.
The variable $\Delta$ is the time delay between the detector and the center of the Earth.
For long signals, this time delay implicitly depends on sidereal time and the location of the source due to the rotation of the Earth.

We Fourier transform the product in Eq.~\ref{eq:template} to obtain an expression in the frequency domain where we do our likelihood evaluations because the covariance matrix can be approximated as diagonal. Therefore, the frequency domain expression becomes:
\begin{align}
    h(f) = & \sum_A \int dt \, F_A\big(t\big) h_A\big(t\big) e^{i2\pi ft}, \label{eq:fourier_transform}\\ 
    = & \sum_A \int dt \, F_A\big(t\big)A_A\big(t\big) e^{i\big(2\pi ft-\phi(t)\big)}. \label{eq:template_f}
\end{align}
Here, we have made use of the post-Newtonian approximation to decompose the strain polarisations into their amplitude $A_{+/\times}\big(t\big)$ and phase $\phi\big(t\big)$. Then, we apply the stationary phase approximation, which is valid provided the amplitude of the time domain signal varies slowly with respect to the phase. Under this scenario, the Fourier transform can be simplified by expanding the integral around the point of stationary phase $t^*$, determined by setting the time derivative of the argument of the exponent in Eq.~\ref{eq:template_f} to zero so that:
\begin{align}
    \label{eq:tstar}
    \frac{d\phi(t)}{dt}\Big|_{t=t^*} &= 2\pi f.
\end{align}
Then, the argument of the exponent can be expanded around $t^*$ in a Taylor series to give 
\begin{align}
    2\pi ft-\phi(t) &\approx 2\pi ft^*-\phi(t^*) - \frac{1}{2}\ddot{\phi}(t^*)(t-t^*)^2 + \dots,
\end{align}
where we ignore terms of higher order than $(t-t^*)^2$. Then, the integral in Eq.~\ref{eq:template_f} is \citep{Cutler_1994}
\begin{align}
    \label{eq:multipole_f}
     h(f) \approx & \sum_A F_A\big(t^*\big)A_A\big(t^*\big) e^{i\big(2\pi ft^*-\phi(t^*)\big)} \nonumber\\
     & \int dt \, e^{-i\frac{1}{2}\ddot{\phi}(t^*)(t-t^*)^2}.
\end{align}
Integrating the Taylor expansion with respect to phase, and substituting the strain polarisations for their post-Newtonian expressions, gives the frequency-domain detector response
\begin{align}
    \label{eq:detectorresponse}
    h(f) &= e^{i\Psi(f)}\sum_AF_A\big(f\big)h_A\big(f\big),
\end{align}
where $F_A(f)\equiv F_A(t^*(f))$ and we have defined the \textit{phase of the detector response} to be
\begin{align}
    \label{eq:phase}
    \Psi(f) &= 2\pi f t^*(f)-\phi\big(t^*(f)\big)-\frac{\pi}{4}.
\end{align}
The stationary point $t^*$ can be defined in terms of frequency through the following relation of the orbital frequency and the energy $E(f)$ and flux $F(f)$ of the binary \citep{Santamaria_2010}
\begin{align}
    \frac{df}{dt} &= -\frac{F(f)}{dE(f)/df},
\end{align}
and is given to lowest order in the post-Newtonian expansion by (setting $G=c=1$) 
\begin{align}
    \label{eq:pnlow}
    t^*(f) &= t_c - \frac{5}{256}\mathcal{M}_c^{-5/3}(\pi f)^{-8/3}.
\end{align}

For a more detailed derivation, we refer the reader to \cite{Iacovelli_2022}.
At low frequencies, incident gravitational wave signals spend hours in the observing band. Over this duration the Earth rotates significantly introducing a frequency dependence in the antenna response factors $F_A$ and phase of the detector response $\Psi$.
At high frequencies, the detector arm length becomes comparable to the wavelength of incident gravitational waves, which also introduces a frequency dependence in the antenna response factors.
Now that the detector response in Eq.~\ref{eq:detectorresponse} includes frequency dependent antenna response factors and phase, we can properly account for the effects due to Earth's rotation and the free spectral range.

Reduced order modelling constructs a sparse representation of the gravitational waveform in two steps: 
\begin{enumerate}
    \item construct a reduced basis out of a training set of waveforms, 
    \item interpolate the space of waveforms using the reduced basis with minimal loss in accuracy.\footnote{In mathematics, ``sparse representation'' refers to the idea of representing signals with some small number of ``atoms'' (component functions).}
\end{enumerate}

\textit{Step 1:} The reduced basis algorithm selects an $N$-component, optimal\footnote{The reduced basis algorithm is considered optimal in the sense that the decay rate of the approximation error of the basis is bounded by the decay rate of the Kolmogorov $N$-width \citep{devore}---the minimum possible approximation error of a manifold $V$ onto an $N$-dimensional subspace $V_N\subset V$.} basis $\{e_i\}$ from a training set of waveforms \citep{rb_pde}. 
Since we are building a reduced order model for gravitational-wave signals from binary neutron stars, each element $e_i$ is an orthonormalised gravitational waveform as measured by an interferometric detector.
In order to achieve this goal, the algorithm goes through iterations.

At each iteration, the algorithm projects the training set waveforms onto the reduced basis. 
The waveform with the worst projection error is added to the basis. 
This minimises the \textit{maximum} approximation error of the basis. 
The algorithm iterates until the maximum approximation error of the basis is less than a user-defined tolerance, which we set to double machine precision $\epsilon=10^{-16}$. 
The relationship between maximum approximation error and the number of basis elements is shown in Fig.~\ref{fig:rb_error}. At $\epsilon = 10^{-14}$, the mismatch---which is defined as a positive-definite quantity---becomes negative due to numerical instabilities and we can no longer reliably compute the error.

The reduced basis method is sometimes preferred over alternative basis construction methods (such as singular value decomposition) because it provides guarantees on the \textit{maximum} approximation error $\epsilon_\text{max}$. 
In the case of gravitational waves, it is important that the error is small compared to the \textit{noise} in a way we quantify below.
If this condition is not met, then the approximation error becomes a systematic error that can affect astrophysical inference.
Thus, it is important to know $\epsilon_\text{max}$ since this determines the maximum signal-to-noise ratio of signals that can be studied without bias.
Methods that do not guarantee $\epsilon_\text{max}$ ``work until they don't''---that is, they sometimes produce significantly biased results, even if they also sometimes produce reliable results.

The empirical interpolation method uses the reduced basis to construct a sparse representation of the waveform in the frequency domain. 
The empirical interpolant is referred to as the \textit{reduced order model}. 
\textit{Reduced order modelling}, meanwhile, refers collectively to the basis construction technique and the use of empirical interpolation. 

\textit{Step 2:} 
The model can be further compressed by replacing the regularly spaced frequency bins (obtained when one Fourier transforms the gravitational-wave strain time series) with $N$ \textit{frequency nodes} $\{F_i\}$. 
There are $N$ such nodes---the same as the number of basis elements. This is because the empirical interpolation algorithm selects a unique node for each basis element to minimise the maximum approximation error of the interpolant. 
The interpolant is sparse when $N$ is less than $L$, the number of data points in the original time series:
\begin{align}
    L = T_\text{m} (f_\text{max} - f_\text{min}) .
\end{align}
Here, $T_\text{m}$ is the duration of the signal/data---which is typically rounded to the next-highest power of two---and $(f_\text{min}, f_\text{max})$ are the minimum and maximum frequencies. In practice, reduced order models in gravitational-wave astronomy are extremely sparse so that $N \ll L$.

The empirical interpolant is \citep{Barrault_2004}
\begin{align}
    \label{eq:ei}
    h^{\text{interp}}(f,\theta) &= \sum_{j=1}^{N} h(F_j,\theta)B_j(f),
\end{align}
where $B_j$ are elements of an interpolation matrix constructed from the reduced basis elements.\footnote{Interestingly, the interpolation matrix and frequency nodes in Eq.~\ref{eq:ei} can be used to construct reduced order model representations for other models that are sufficiently similar to the original model, i.e., other models that are in the span of the reduced basis.}
Here $\theta$ are the binary parameters like component masses, distance to the source, etc.
The elements of the interpolation matrix are defined as 
\begin{align}
    B_j(f) &= \sum_i^N e_i(f)(V^{-1})_{ij},
\end{align}
where $\{e_i\}$ are the orthonormalised reduced basis elements and $V$ is a matrix with elements defined by
\begin{align}
    V_{ij} &= e_i(F_j).
\end{align}
The accuracy of the empirical interpolant is typically $100$ times worse than the maximum reduced basis error for gravitational waveforms \citep{Field}. 
Reduced order models can be used, as in this work, to form a compressed likelihood. However, reduced order modelling can also be used to build \textit{surrogates} for fast waveform evaluation, enabling the use of otherwise prohibitively slow waveforms (see, e.g. \cite{chatziioannou2024compactbinarycoalescencesgravitationalwave}).

\subsection{Reduced Order Quadrature}
The standard log likelihood used for gravitational-wave parameter estimation is
\begin{align}
    \label{eq:loglikelihood}
    \ln \mathcal{L} = \langle d,h \rangle - \frac{1}{2}\langle h,h\rangle -\frac{1}{2}\mathcal{Z}_n,
\end{align}
where $d$ is the data, $h$ is a template waveform, and $\mathcal{Z}_n\equiv\langle d,d\rangle$ is the noise evidence.
Following the notation from \cite{thrane_talbot_2019}, $\langle a, b \rangle$ is the noise-weighted inner product of $a$ with $b$, which includes a sum over evenly-spaced frequency bins:
\begin{align}\label{eq:inner_product}
    \langle a , b \rangle \equiv \frac{4}{T} \sum_{j=1}^L \frac{\Re (a_j^* b_j)}{S_h(f_j)}
\end{align}
Here, $S_h(f)$ is the single-sided strain noise power spectral density.

Evaluating the sum in Eq.~\ref{eq:inner_product} is computationally expensive. 
By replacing the waveform $h$ with the reduced order model in Eq.~\ref{eq:ei}, one can instead sum over sparse elements.
The resulting expression is referred to as the \textit{reduced order quadrature} (ROQ) likelihood: 
\begin{align}
    \label{eq:roq}
    \ln \mathcal{L} = &\sum_{j=1}^{N_L} h(F_j,\theta) \omega_j -\frac{1}{2}\sum_{i=1}^{N_Q}|h(F_i,\theta)|^2\psi_{i} - \frac{1}{2}\mathcal{Z}_n.
\end{align}
The quantities $\omega_j$ and $\psi_{i}$ are integration weights that can be computed in pre-processing. 
The variables $N_L$ and $N_Q$ are the number of basis elements required for the reduced order model of the \textit{Linear} and \textit{Quadratic} detector response terms in Eq.~\ref{eq:roq}.

In this work we build a reduced order model that can be compared directly with the data as in Eq.~\ref{eq:template}.
This is contrast to previous work by \cite{Smith_2016, rory, Morisaki_2023}, which built models for $h(t)$ ignoring the effects due to the free spectral range and only including the change in detector orientation due to Earth's rotation. These studies ignore the time dependent time delay $\Delta(t)$ that arises from the change in the light travel time from the Earth's geocenter to the detector location as Earth rotates.  
This allows the phase to be absorbed into the weights resulting in smaller ROMs, but requires building the weights for $\mathcal{O}(10^5)$ parameter values to interpolate between. 
Instead, we incorporate the phase directly into the empirical interpolant.
This is necessary to include the changing light travel time and therefore fully include the effects of Earth's rotation. The phase becomes time-dependent and rapidly-varying, and can no longer be straightforwardly absorbed into the weights. This means we no longer need to interpolate the weights which greatly reduces the startup cost of reduced order inference.

For a typical binary neutron star signal in Cosmic Explorer ignoring Earth's rotation and the free spectral range effects, only a few hundred basis elements are required and the likelihood speedup using ROQ is 
\begin{align}
    \label{eq:speedup}
    \frac{L}{N_L+N_Q} \sim \mathcal{O}(10^4).
\end{align}
For details on the derivation of the reduced order quadrature likelihood in Eq.~\ref{eq:roq} and how it differs to those used in other works, we refer the reader to Appendix~\ref{appendix:roq_choice}.

\subsection{Accuracy of ROQ}\label{sec:accuracy_of_roq}
Having covered how to construct a reduced order likelihood, we now discuss how to determine if the reduced order approximations are adequate for inference calculations at a given signal-to-noise ratio.
To do so, we employ an approximate relation between the relative error in the log likelihood ratio  and the maximum signal-to-noise ratio $\rho$ able to be studied by a reduced order quadrature rule:
\begin{align}\label{eq:delta_ll_snr}
    \frac{\Delta \ln \mathcal{L}}{\ln \mathcal{L}} &\lesssim \frac{1}{\rho^2}.
\end{align}
The likelihood error is defined as 
\begin{align}
    \Delta\ln\mathcal{L} = \ln\mathcal{L} - \ln\mathcal{L}_{\text{ROQ}} ,    
\end{align}
where $\ln\mathcal{L}$ is the standard likelihood and $\ln\mathcal{L}_{\text{ROQ}}$ is the ROQ likelihood. 
This rule is demonstrated in Appendix~\ref{appendix:aprior}.  

\subsection{Constructing the Training Set}\label{sec:construc_training_set}
Our model is constructed using the  \texttt{IMRPhenomPv2\_NRTidalv2} approximant. 
Millions of waveform evaluations are necessary to construct an accurate model.
We utilise a number of strategies in order to control the size of the training set: $(1)$ adaptively sample the frequency domain, $(2)$ iteratively enrich the training set, and $(3)$ utilise targeted sampling techniques and a noise-weighted inner product to build the reduced basis.

\textit{Adaptive sampling.} The first strategy we use to minimise the size in memory of the training set is evaluating waveforms over adaptively sampled frequency domains. The steps are outlined below:

\begin{itemize}
    \item Split the frequency domain into bands $b$ with maximum frequencies according to $2^n$: $\unit[5]{Hz}, \unit[8]{Hz}, \unit[16]{Hz}, $ etc.
    Since the inspiral signal is quasi monochromatic, it is a good approximation to say that the signal is in each band for only a fraction of the total signal duration.
    This allows us to downsample each band according to the amount of time that the signal is present in each band.
    Down-sampling reduces the number of frequency bins, which makes the calculation more tractable.
    \item downsample each band with an adaptive resolution $\Delta f^{(b)}$ based on the time spent in each band by the lowest mass binary in parameter space. 
    Thus, for example, the 5-$\unit[8]{Hz}$ band has a resolution of $\Delta f = 1.2\times10^{-4}$ while the 32-$\unit[64]{Hz}$ band has a resolution of $\Delta f = 1.5\times10^{-2}$.
    The 5-$\unit[8]{Hz}$ band requires higher resolution than the 32-$\unit[64]{Hz}$ band because the binary spends $128$ times longer in the former than the latter.
    \item Build the reduced basis from a training set of downsampled waveforms. 
    \item Take the (few) parameters $\{\theta_i\}_i^{N_L}$ that give the downsampled basis elements. Build a training set of waveforms evaluated over the full frequency array $L$ at these parameters. We now have a faithful training set. As this training set is only a few hundred faithful waveforms, it is relatively inexpensive compared to directly constructing a faithful training set of millions of waveforms.
    \item Build the faithful reduced basis over this new training set. Because the downsampled waveforms were sampled above the Nyquist rate, we can safely assume the parameters identified by the downsampled reduced basis accurately represent the most informative waveforms in the faithful case. We validate this by computing the mismatch between the reduced order model and the true waveform for $50,000$ parameter values.
\end{itemize}
This adaptive sampling technique reduces the memory requirements of the training set by $\mathcal{O}(10^2)$ and enables fast construction of faithful bases.

The time in each band is determined using the post-Newtonian expansion of the time to merger taken to second order, which is an adequate approximation for the early inspiral:
\begin{align}\label{eq:Tm}
    T_\text{m} =& \frac{5}{256}\mathcal{M}_c^{-5/3}(\pi f)^{-8/3} \nonumber\\
    \propto & f^{-8/3} ,
\end{align}
where $\mathcal{M}_c$ is the chirp mass.
The post-Newtonian approximation breaks down near the inner-most stable circular orbit. Therefore, we limit the maximum frequency resolution to $\Delta f^{(b)} = 1$ to ensure that we do not under-sample the high-frequency bands. 
To avoid sampling below the Nyquist rate, the time spent in each band is rounded to the nearest power of two. 
When building the reduced basis over the faithful training set, we track the projection errors to validate the adaptive sampling procedure. 

\textit{Iterative enrichment.} The second technique we employ to minimise the size of the training set is iterative training set enrichment. 
The basic idea is to validate our model over multiple training sets. 
It is unlikely that two randomly-constructed training sets will have the same waveforms, so this effectively increases the number of training set waveforms we build our model over.
We build three supplementary training sets, which is enough to ensure model accuracy. 
Each training set is projected onto the reduced basis, and any waveform with approximation errors greater than the user-defined tolerance $\epsilon$ is added to the basis.

\textit{Targeted sampling.} Finally, we draw waveforms from regions of parameter space that we know are important for accurate models.
We draw half of our waveforms randomly from the standard gravitational-wave prior.\footnote{We set uniform priors on all parameters. The chirp mass prior is given by $1.18\leq\mathcal{M}_c\leq 1.18+5\Delta\mathcal{M}_c$, where $\Delta\mathcal{M}_c=8\times 10^{-3}\times (32/\text{SNR})$. We limit the mass ratio to $0.5\leq q \leq 1$, the dimensionless spin magnitudes to $|a_{1/2}|\leq 0.05$, the luminosity distance to $1\leq d_L\leq\unit[4000]{Mpc}$, the tidal deformabilities to $0\leq\Lambda_{1/2}\leq 1000$, and the time of coalescence to $|t_c|\leq \unit[0.1]{s}$. We use the full range for all other parameters.}
Following \cite{rory}, the other half are drawn from a regularly spaced grid. The grid is uniform in all parameters other than chirp mass, which we sample uniformly in $\mathcal{M}_c^{-5/3}$ as this term describes the inspiral phase at leading order. There are 500 grid points in each dimension. We draw randomly from the multi-dimensional grid. Drawing points from a grid ensures that half of the training set waveforms are uniformly spaced, which ensures a more even sampling of the parameter space.

It has also been shown that basis elements are preferentially selected from the boundary of parameter space \citep{Smith_2016}. 
This may arise from the slowly varying and smooth nature of the space of waveforms, such that the most distinct waveforms lie on opposite ends of the parameter space boundary.
Therefore, all boundary parameter samples are included in the training set. 
We also construct the reduced basis using the noise-weighted inner product. 
Then, the greedy algorithm selects basis elements that are more informative with respect to the likelihood, further decreasing the size of the basis.

\subsection{Earth's Rotation and the Free Spectral Range}
Over the hours-long signals expected to be observed by next-gen detectors, the detector moves through space as Earth rotates. 
Furthermore, as the arm length of the detector $L_\text{arm}$ approaches the wavelength of the incident wave, the signal evolves over the travel time of photons in the detector arms.
The frequency at which this occurs is the free spectral range (FSR), given by
\begin{align}
    f_{\text{FSR}}=\frac{c}{2L_\text{arm}}.
\end{align}
Here, $c$ is the speed of light.
These two phenomena---the rotation of the Earth and the FSR---make the antenna response functions $F^{+/\times}$ and the detector response phase $\Psi$ (Eq.~\ref{eq:phase}) frequency-dependent.
The antenna response is given by \citep{Essick_2017}
\begin{align}\label{eq:fpc}
    F^{+,\times}(\hat{n}|t) &= D_{ij}(\hat{n} | t) \, \epsilon^{ij}_{+,\times}(\hat{n}),
\end{align}
where $D_{ij}$ is the detector tensor, $\epsilon^{ij}_{ +,\times}$ is the polarisation tensor, $\hat{n}$ is the direction to the gravitational-wave source, and $t$ is time.\footnote{The indices $i$ and $j$ run over the three spatial dimensions while $+$ and $\times$ describe the two polarization states of gravitational waves. Here, the repeated indices imply summation.}
In the frequency domain, Eq.~\ref{eq:fpc} becomes:
\begin{align}
    F^{+,\times}(\hat{n}|f) = D_{ij}(\hat{n} | f) \, 
    \epsilon_{+,\times}^{ij}(\hat{n}) .
\end{align}
Incorporating the effects due to Earth's rotation and the free spectral range causes both the magnitude and phase of $D_{ij}(f)$ to vary with frequency.
This in turn means that the amplitude and phase of $F^{+,\times}(\hat{n} | f)$ to vary with frequency.
This is in contrast to analyses carried out on LIGO/Virgo data where one can safely approximate $F^{+,\times}(\hat{n} | f)$ as a constant in frequency.
This frequency dependence is illustrated in Fig.~\ref{fig:fp_phase_vs_constant} of Appendix~\ref{app:plots}, where we compare the frequency-dependent and static antenna response and phase.

The reduced basis algorithm selects basis elements based on unique features. 
The varying phasing of the detector response introduces complexity in the structure of the waveforms. 
For one thing, the frequency dependence is different for different directions $\hat{n}$.
The rapidly varying structure of the waveforms cause the basis to become extremely large, making it impractical with current computational resources to construct accurate ROMs over full parameter domains for $f_{\min}\leq \unit[17]{Hz}$. 
As a result, \textit{there is currently no technique in the literature} that provides fast and accurate inference for next-gen gravitational wave detectors. 
We demonstrate this in detail in Section~\ref{sec:methodology}, where we construct reduced order models across all 17 parameters that parameterise a binary neutron star waveform.

\section{Alternatives to reduced order modelling}\label{sec:alternatives}
Before we demonstrate the limitations of reduced order modelling, we pause to comment on other methods that have been proposed to control the computational cost of inference.
The most notable alternatives to reduced order modelling are multi-banding \citep{multibanding} and relative binning \citep{zackay2018relative, modebymode_relativebinning}. 
Perhaps relative binning and multi-banding could be used to make inference in next-generation detectors, however we are worried that these techniques cannot achieve better compression than reduced order modelling without sacrificing accuracy.
Qualitatively, our argument can be understood as follows.
Reduced order methods, multi-banding, and relative binning are all methods of data compression.
However, the complexity arising from the FSR and the rotation of the Earth significantly appears to be too much for compression schemes to overcome.
There is a finite amount of compression possible based on the variability of waveforms in our signal space.
Reduced order modelling appears to provide the most efficient compression scheme, with a guaranteed minimum error. 
Other compression schemes cannot overcome this added complexity without sacrificing precision.
As always, there is no free lunch.

We demonstrate this quantitatively in Appendix~\ref{appendix:aprior}.
In Fig.~\ref{fig:rb_roq_rel_ll_error} we show that relative binning and multi-banding likelihoods constructed without the detector effects described in this Section are orders of magnitude less accurate than the reduced order quadrature likelihood that incorporates these effects. We use the \texttt{Bilby} implementation of the relative binning \citep{krishna2023acceleratedparameterestimationbilby} and multi-banding likelihoods in our comparison.
We expect the gap in accuracy and acceleration between these techniques to widen when incorporating detector effects into multi-banding and relative binning, as well as lowering $f_{\min}$. In addition, reduced order quadrature is preferable as it provides \textit{a priori} guarantees on the maximum signal-to-noise ratio that can be studied without introducing bias in the inference.

\section{Reduced order modelling with limited memory }\label{sec:methodology}
In this section we show that reduced order modelling---using currently available computers---is not a practical solution for the loudest binary neutron star events observed with next-gen observatories.
It is, in fact, extremely challenging to analyse a single event like GW170817 without throwing away data below $\approx\unit[16]{Hz}$ in order to make the calculation tractable.

We construct a set of reduced order models for a binary neutron star signal with different minimum frequencies $f_\text{min}$.
We require that each model is no more than $\unit[100]{Gb}$ in size, and that it can be stored using variables of double precision.
Anything larger than $\unit[100]{Gb}$ becomes unwieldy.
By limiting the size of our models to $\unit[100]{Gb}$, we limit the number of basis functions $N$, which in turn limits the accuracy of the model.
The accuracy of the model, meanwhile, determines the maximum SNR that we can analyse without obtaining biased results via Eq.~\ref{eq:delta_ll_snr}.
Thus, the $\unit[100]{Gb}$ imposes an indirect limit on the maximum SNR.

\begin{figure*}[h]
    \centering
    \includegraphics[width=\linewidth]{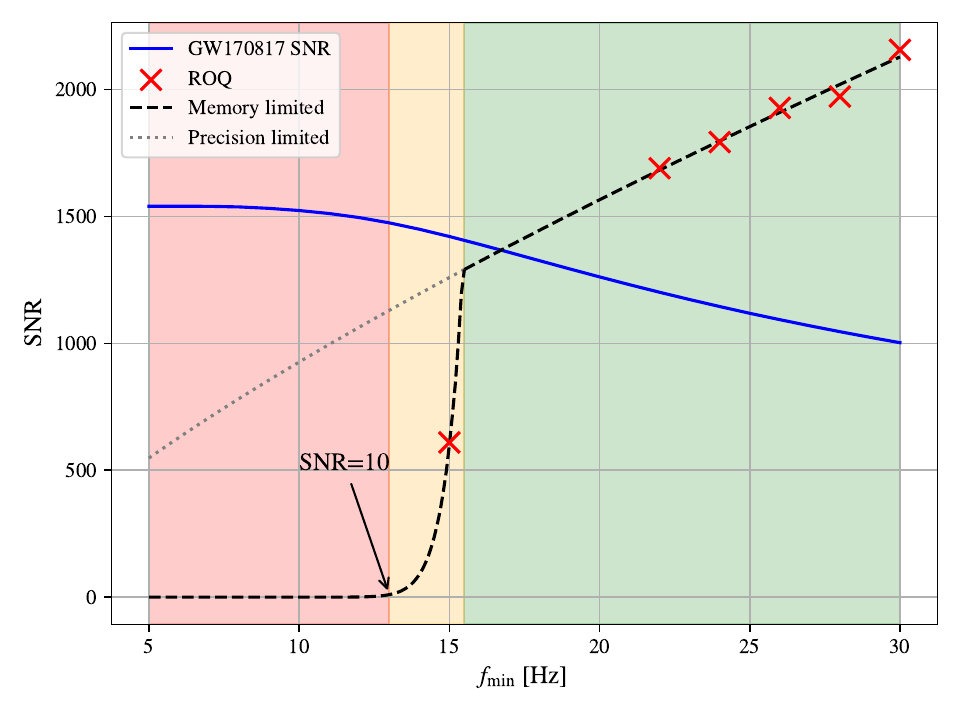}
    \caption{
     The limitations of inference with data compression strategies.
     Solid blue shows the SNR of a GW170817-like event as observed by a single Cosmic Explorer. 
     This is contrasted with the dashed black showing the maximum signal-to-noise ratio that that we can currently achieve with reduced order modelling. 
     In the green region above $\unit[16]{Hz}$, the maximum SNR is limited only by double machine precision, which is almost sufficient.
     In the orange region between $\unit[13-16]{Hz}$, the maximum SNR is limited by our memory limit of $\unit[100]{Gb}$; we can carry out inference, but only by artificially inflating error bars.
     In the red region below $\unit[13]{Hz}$, the maximum SNR drops below 10, and we are unable to carry out meaningful calculations, even by inflating error bars.
     }
    \label{fig:snr_curve}
\end{figure*}

As we decrease $f_\text{min}$, the signals spend more time in band; the time to merge scales according to Eq.~\ref{eq:Tm}.
Moreover, as we decrease $f_\text{min}$ the signal morphology becomes more complicated due to the rotation of the Earth.
Thus, each basis element becomes larger and more basis elements are required to achieve a fixed accuracy.
As a result, the maximum SNR decreases with $f_\text{min}$. 
Our goal is to determine the relationship between maximum SNR and $f_\text{min}$.
We compare the maximum SNR with the expected SNR of a GW170817-like binary neutron star with masses $m_{1,2} = 1.4 M_\odot$ at a distance of $\unit[40]{Mpc}$.
For a reduced order model to work, we need the maximum SNR to exceed the SNR from a GW170817-like event.
We study minimum frequencies between 5-$\unit[30]{Hz}$.
Over this range, the expected SNR of a GW170817-like event varies from $\text{SNR}\approx 1600$ at $\unit[5]{Hz}$ to $\text{SNR}\approx 1000$ at $\unit[30]{Hz}$; see the solid blue curve in Fig.~\ref{fig:snr_curve}.

The first step is to determine the range of chirp mass $\Delta M_c$ over which the reduced order model should be built for a given value of $f_\text{min}$.
We want $\Delta M_c$ to be as small as possible in order to minimise the size of the model.
However, it must be large enough to safely include the tails of the posterior distribution.
We choose $\Delta M_c$ to include $\pm 5$ standard deviations from the true value of $M_c$.\footnote{Of course, in an actual analysis, the true value of $M_c$ is unknown, but a maximum-likelihood estimate is provided from a matched filter search upstream of the parameter estimation analysis.}
The size of one standard deviation depends primarily on the signal-to-noise ratio \citep{Cutler_1994}, so we estimate $\Delta M_c$ with the scaling relation:
\begin{align}\label{eq:scaling}
    \Delta M_c \propto \frac{1}{\text{SNR}} .
\end{align}
We know $\Delta M_c = \unit[8.3\times10^{-3}]{M_\odot}$ for GW170817, which was detected with SNR=32; we use this data point to estimate $\Delta M_c$ for other SNR values.

The next step is to build the reduced order model with sufficient accuracy to analyse the GW170817-like event at the expected SNR given by the blue curve in Fig.~\ref{fig:snr_curve}.
We add basis functions until either (1) we succeed and the accuracy is limited by double machine precision or (2) we fail because the model exceeds $\unit[100]{Gb}$.

For $f_\text{min}>\unit[16]{Hz}$, the procedure is a success and we construct models limited by machine precision, which are less than $\unit[100]{Gb}$ in size. 
The relationship between maximum SNR and $f_\text{min}$ above $\unit[16]{Hz}$ is marked on Fig.~\ref{fig:snr_curve}: red crosses show reduced order models we have successfully constructed while the black dashed line shows the empirical scaling relation,
\begin{align}\label{eq:snr_vs_fmin}
    \text{SNR}_\text{max} \propto f_\text{min}^{0.71} .
\end{align}
This scaling relation will prove useful momentarily as we seek to extrapolate to lower values of $f_\text{min}$ where it is more computationally challenging to construct a reduced order model.

Between 16-$\unit[17]{Hz}$, the procedure succeeds in staying under the memory requirement, but the model falls just short of the required accuracy due to the machine precision of our double arrays.
This causes the dashed maximum SNR curve in Fig.~\ref{fig:snr_curve} to fall slightly below the solid blue curve showing the SNR of a GW170817-like event.
However, this machine precision limitation is a relatively minor problem compared to the problems that arise below $\unit[16]{Hz}$.
Below this $f_\text{min}$, the analysis becomes limited by memory.

Below $\unit[16]{Hz}$, we reduce the SNR by artificially inflating the detector noise.
However, reducing the SNR necessitates that we broaden the chirp mass prior according to Eq.~\ref{eq:scaling}.
This leads to a chicken-and-egg problem: we need the maximum SNR to know how wide to make $\Delta M_c$ so we can calculate the maximum SNR.
Fortunately, we can use empirical scaling laws to determine $\Delta M_c$ and SNR.

The first scaling relation that we need is one between the number of basis elements $B_\text{max}$---required for a model rated up to $\text{SNR}_\text{max}$---and the minimum frequency:
\begin{align}\label{eq:basis_size_fmin}
    B_\text{max} \propto f_{\min}^{-2.2} .
\end{align}
Together, Eq.~\ref{eq:snr_vs_fmin} and Eq.~\ref{eq:basis_size_fmin} imply that
\begin{align}\label{eq:max_max}
    \text{SNR}_\text{max} \propto B_\text{max}^{0.32} .
\end{align}
However, note: this is only true when we hold fixed the error tolerance $\epsilon = 10^{-16}$ as required by machine precision.

Next we need an empirical relationship between $\text{SNR}$ and $B$ for \textit{variable} error tolerance.
We construct models for three values of $f_{\min}$: $\unit[26]{Hz}$, $\unit[28]{Hz}$, and $\unit[30]{Hz}$.
For each $f_\text{min}$, we construct models for SNR$=\{250, 500, 750, 1000\}$ using variable error tolerances $\epsilon$ (and thus variable chirp mass ranges $\Delta M_c$). 
For each of these $f_{\min}$ we fit the relation between SNR and basis size assuming a power law. 
We obtain the following relation: 
\begin{align}\label{eq:snr_basis_fmin_plaw}
    \text{SNR} =
    \kappa(f_\text{min})
    B^{8.3-2.0(f_{\min} / \unit[10]{Hz})}.
\end{align}
Here, $\kappa$ depends on $f_\text{min}$, but---since we are about to apply this equation at a fixed $f_\text{min}$---$\kappa$ is a constant for our purposes.

The final thing we need to do before we can use these scaling relations to calculate the relationship between $\text{SNR}$ and $f_\text{min}$ is to relate the number of basis elements $B$---and the size of each element---to the \textit{size} of the basis---which we require to be less than $\unit[100]{Gb}$.
The size of the basis in bytes is 
\begin{align}\label{eq:basis_size_gb}
    \mathcal{S} = & 16 B T_\text{m} (f_{\max}-f_{\min}) .
\end{align}
Here $f_{\max}-f_{\min}$ is the bandwidth (we take $f_{\max}=\unit[2048]{Hz}$), $16$ is the number of bytes in each double-precision variable used to represent the waveform, and the time to merge is given in Eq.~\ref{eq:Tm}. 
Our requirement that $\mathcal{S} < \unit[100]{Gb}$ implies a maximum value for $B$ given by 
\begin{align}\label{eq:basis_100gb}
    B_{100} = \frac{10^{11}}{16 T_\text{m} (f_{\max}-f_{\min})} .
\end{align}

Finally, we are ready to calculate the memory-limited signal-to-noise ratio, denoted $\text{SNR}_{100}$, as a function of minimum frequency for $f_\text{min}<\unit[16]{Hz}$---the region where the maximum SNR model exceeds $\unit[100]{Gb}$.
The steps are as follows:
\begin{enumerate}
   \item Use the scaling relation in Eq.~\ref{eq:snr_basis_fmin_plaw} to relate $(B_{100}, \text{SNR}_{100})$ to $(B_\text{max}, \text{SNR}_\text{max})$. This yields:
    \begin{align}\label{eq:SNR100}
        \text{SNR}_{100} &= \left(\frac{B_{100}}{B_{\max}}\right)^{8.3-2.0(f_\text{min}/\unit[10]{Hz})}\text{SNR}_\text{max}.
    \end{align}
    \item Determine $B_{100}$: the number of basis elements required for a $\unit[100]{Gb}$ basis using Eq.~\ref{eq:basis_100gb}. 
    \item Determine $\text{SNR}_\text{max}$: the maximum signal-to-noise ratio allowed by machine precision using Eq.~\ref{eq:snr_vs_fmin}.
    \item Determine $B_\text{max}$: the number of basis elements associated with $\text{SNR}_\text{max}$ using Eq.~\ref{eq:basis_size_fmin}.
\end{enumerate}

Putting everything together, yields the following scaling relation for $f_\text{min}<\unit[16]{Hz}$:
\begin{align}\label{eq:snr100_fmin}
    \text{SNR}_{100} &\propto \left(f_{\text{min}}\right)^{40 -9.8 (f_\text{min}/\unit[10]{Hz})}\text{SNR}_\text{max}(f_\text{min})
\end{align}
This scaling leads to the precipitous drop in the dashed black curve below $\unit[16]{Hz}$ in Fig.~\ref{fig:snr_curve}. 
The black dashed curve can be contrasted with the dotted grey curve, which shows the $\text{SNR}_\text{max}$ versus $f_{\text{min}}$ if we had no $\unit[100]{Gb}$ limit so that the models are limited only by machine precision. 

To validate these calculations, we build a ROM for $f_{\text{min}}=\unit[15]{Hz}$ and determine the maximum $\text{SNR}$ to be $\text{SNR}=608$ which matches the theoretical prediction of $\text{SNR}=601$. The size of the basis is $\unit[33]{Gb}$ in memory, which is relatively close to our $\unit[100]{Gb}$ limit. 
Therefore our scaling relations predict the memory to within a factor of a few, which is adequate for the purposes of demonstrating a problem that is many orders of magnitude larger.
This $\unit[15]{Hz}$ ROM is indicated by a red cross in Fig.~\ref{fig:snr_curve}.

We divide Fig.~\ref{fig:snr_curve} into three colored regions.
In the green region above $f_\text{min}=\unit[16]{Hz}$, we can successfully build models that with sufficient accuracy to analyze GW170817-like events.
In the yellow region between $f_\text{min}=13$-$\unit[16]{Hz}$, we can construct models with $\text{SNR}>10$ by artificially inflating the noise.
This is undesirable because we are throwing away information, but it shows that the calculation is at least close to tractable.
In the red $f_\text{min}<\unit[13]{Hz}$ region, the SNR falls below 10, and it becomes difficult to construct useful models.
A network SNR of 12 is typically required for an event to be considered a detection.
Even for a single event, it is therefore challenging to carry out inference down to these $f_\text{min}$ values using contemporary computers and with existing data compression strategies.

\section{Inference on a Binary Neutron Star Signal}\label{sec:inference}
In this section we demonstrate inference on a binary neutron star signal like GW170817, taking into account the rotation of the Earth and the FSR.
Previous efforts to accurately perform inference on a binary neutron star signal observed by a next-generation detector missed key details. 
\cite{rory} ignores the effects due to the free spectral range, which contributes to the prohibitive computational expense in construction ROMs. 
\cite{baral} utilises multibanding \citep{multibanding} and the \texttt{TaylorF2} waveform, neither of which are guaranteed to be accurate for an $\text{SNR}=1000$ binary neutron star siganl. 
To our knowledge, this is the first demonstration of inference on an $\gtrsim\unit[1]{hr}$-long binary signal, done using a reduced order model in order to guarantee the likelihood is sufficiently accurate for reliable inference, and including the effects of the rotation of the Earth and the FSR.

We use the reduced order model starting from $f_{\min}=\unit[26]{Hz}$ (safely in the Fig.~\ref{fig:snr_curve} green zone) with a maximum frequency of $\unit[2048]{Hz}$ and a signal duration of $\unit[128]{s}$. 
The linear and quadratic reduced bases have $N_L = 2667$ and $N_Q = 2425$ elements, respectively, giving a theoretical likelihood acceleration of 50 (Eq.~\ref{eq:speedup}). 
We inject a binary neutron star signal generated using the \texttt{IMRPhenomPv2\_NRTidalv2} approximant \citep{Dietrich:2019kaq} at $d_L=\unit[40]{Mpc}$ into a single Cosmic Explorer located at the site of the current LIGO Hanford observatory with a zero-noise realisation of Gaussian noise. 
The injection step includes effects from the rotation of the Earth and the FSR.
This yields $\text{SNR} = 792$, which is well within the accuracy bounds of the ROQ rule. 

We assume standard priors on the 17 parameters that characterise a binary neutron star merger. We utilise uniform priors on chirp mass $\mathcal{M}_c$, mass ratio $q$, spin magnitudes $\chi_{1,2}$, and the tidal deformabilities $\Lambda_{1,2}$. 
We adopt an isotropic prior for the direction of the spin vectors.
Furthermore, we assume a prior that is uniform in comoving volume and source frame time for the luminosity distance $d_L$.
We adopt standard priors for the other extrinsic parameters.
By injecting into zero noise, we expect the posterior distribution to peak at the injected parameter values. This serves as a useful diagnostic to verify the accuracy of the ROQ likelihood in Eq.~\ref{eq:roq} and the ROMs constructed in this work.

We use the \texttt{dynesty} \citep{dynesty} nested sampling package as implemented within \texttt{Bilby} \citep{bilby,isobel_2021} to estimate the posterior distribution. We use $2000$ live points with the acceptance walk Markov chain Monte Carlo method from \cite{bilbymcmc}. The sampling is parallelised over 124 processes. The inference takes 2.3 hours (285 CPU hours) during which the log likelihood is called 50M times. 
The average log likelihood evaluation time is $\unit[0.25]{s}$. 
Therefore, the total sampling time without using reduced order quadrature would be approximately 3527 CPU hours, or 28.4 hours using 124 processes. 
The reduced order quadrature likelihood accelerates inference from 3527 CPU hours to 285 CPU hours, giving an empirical inference speedup of a factor of $12$---less than the theoretical prediction of $50$ due to overheads in the inference procedure.

In Figs.~\ref{fig:26hz_masses_spins}--\ref{fig:26hz_tides_time}, we show in blue the one and two dimensional posterior distributions for various combinations of parameters. 
The posterior distributions peak at the true parameter values, indicated in orange, validating the reduced order inference procedure outlined in this work. 
The location of the source in the sky is constrained to $\unit[403]{deg}^2$ at the $90\%$ credibility. 
This sky area is far larger than the field of view of the upcoming Vera Rubin observatory, planned to be used for electromagnetic follow-up observations of binary neutron star mergers \cite{Chen_2021}. 
Narrow constraints on sky localisation are crucial for rapid multi-messenger observations. 
Whilst we expect that the constraints on sky location will improve significantly when considering signals beginning from $f_{\min}=\unit[5]{Hz}$, we face severe limitations in achieving fast and accurate inference below $f_{\min}=\unit[17]{Hz}$.

\section{Conclusion}\label{sec:conclusions}
Next-gen gravitational wave detectors will observe thousands of binary neutron star signals, some of which will be in the detecting band for $\gtrsim\unit[1]{hr}$.
However, after taking into account subtle effects from the rotation of the Earth and the free spectral range, we show that it is challenging to analyze even one of these long signals with reduced order modelling, and we are concerned that no current data analysis strategies can be used on contemporary computers.

Our findings exacerbate computational issues pointed out in \cite{Hu}, who estimate the number of cores required to process the hundreds of thousands of compact binary signals detected by next-gen observatories.
Even using data compression techniques like reduced order modelling, but assuming no paradigm-changing new algorithms, \cite{Hu} find that $10^9-10^{15}$ contemporary CPU cores are likely required to analyse one month of data.
While \cite{Hu} focus on the sheer volume of binary signals---and we agree with their assessment that the large volume of detections is problematic---here we show that it is currently challenging to analyse even just \textit{one} of the loudest binary neutron stars that will be detected by Cosmic Explorer and the Einstein Telescope.

In addition to the raw computational challenges posed by next-gen observatories, the community faces formidable challenges relating to systematic error / model misspecification \citep{PhysRevResearch.2.023151, wmf}.
Key systematic errors include systematic errors in gravitational waveforms, imperfections in the noise model, and calibration error.
While the past decade has seen great strides in the development of fast gravitational waveform approximants, the current level of systematic error---which we can estimate by looking at the differences between different approximants---is large compared to the statistical uncertainty with which the loudest binary signals will be observed by next-gen observatories \citep{Owen}.

Likewise, our ability to accurately describe detector noise is limited by the non-Gaussian character of real interferometer noise \citep[see, e.g.,][]{ligo_noise}. 
And even Gaussian noise models are complicated by subtle effects, which are often ignored in LIGO/Virgo analyses, but which will be significant systematic errors for next-gen observatories \citep{student-t,Biscoveanu_2020,windows,Maximum_entropy,glitch_mitigations2,PSD_effect_on_PE_1,tPowerBilby}.
Researchers have proposed elegant solutions to marginalise over uncertainty in the noise model and to take into account edge effects, but these solutions tend to increase the computational cost of the analysis, exacerbating the problems identified here.

Finally, the calibration uncertainty is expected to be a significant source of systematic error for next-gen observatories~\citep{essick_2022}.
Our point is that---even if we are able to solve the computational problems that make it a challenge to carry out inference on long gravitational-wave signals---it may be necessary to significantly inflate error bars due to systematic errors.

We end on an optimistic note.
First, some of the challenges associated with next-gen observatories seem to be under control.
While LIGO/Virgo detect gravitational-wave signals that are clearly separated in time, Cosmic Explorer and the Einstein telescope will typically observe multiple binary neutron stars at the same time \citep{GW170817_stoch}.
However, results by \cite{Johnson} indicate that the presence of neighbouring signals is unlikely to interfere with astrophysical inference.
And work by \cite{cosmo} suggests it is possible to measure a primordial gravitational-wave background obscured by a foreground of compact binary signals.

Second, while we are pessimistic about the possibility of carrying out inference on long gravitational-wave signals with the current paradigm, we think it may be possible with the development of new techniques.
We argue here that the calculations have reached the maximum benefit possible by compression.
However, significant speed-ups may be possible through deep learning schemes \citep[e.g., ][]{dingo}, which perform amortized inference without the need for likelihood evaluations. 
In fact, recent work has demonstrated the capabilities of deep learning methods to provide real time inference for binary neutron star signals in current detector networks \citep{dax2024realtimegravitationalwaveinferencebinary}.
Additionally, a machine learning based workflow has been suggested for the efficient analysis of long-duration binary neutron star signals in next-generation detectors \citep{hu2024decodinglongdurationgravitationalwaves}. 
Finally, clever parallelisation of the analysis may make it possible to work with manageable-sized reduced order models.
We leave this for future work.

\section{Acknowledgements}
This is LIGO Document number LIGO-P2500070.
This material is based upon work supported by NSF’s LIGO Laboratory which is a major facility fully funded by the National Science Foundation. This work is supported through Australian Research Council (ARC) Centres of Excellence CE170100004, CE230100016, Discovery Projects DP220101610 and DP230103088, and LIEF Project LE210100002. This work was performed on the OzSTAR national facility at Swinburne University of Technology. The OzSTAR program receives funding in part from the Astronomy National Collaborative Research Infrastructure Strategy (NCRIS) allocation provided by the Australian Government, and from the Victorian Higher Education State Investment Fund (VHESIF) provided by the Victorian Government.

\appendix

\section{Building reduced order models with \texttt{Rombus}}\label{appendix:rombus}
\texttt{Rombus} is a publicly available python package for building reduced order models for arbitrary functions. In this section we describe how to make ROMs of gravitational waveforms using \texttt{Rombus}. However, the steps outlined offer a generalised guide for building ROMs of any model. We also describe changes made to \texttt{Rombus} to allow for the efficiency construction of ROMs. Finally, we include descriptions on the use and implementation of a number of gravitational wave specific algorithms utilised throughout this work, which are contained in a \texttt{utils.py} file. This code is publicly accessible here \url{https://github.com/Makai-Baker/ROM_pipeline}, and the version of \texttt{Rombus} used in this work is available here \url{https://github.com/Makai-Baker/Rombus_3G_GW_model_adaptive_sampling}.

The first step in building a ROM is to define the base model to be compressed, which must be saved in a \texttt{.py} file. This model must be a subclass of the generic \texttt{RombusModel} class. Each model must define a domain, codomain,\footnote{In mathematics, a codomain is a set that includes all the possible values of a function---in this case, a set of predicted strain values.} and prior ranges for all model parameters. In this work, the domain is frequency, the codomain is the strain $h$, and the prior ranges are built over the 15 parameters that  parameterise the \texttt{IMRPhenomPv2\_NRTidalv2} waveform approximant. We improve the flexibility of \texttt{Rombus} by allowing arbitrary keyword arguments to be passed to the user-defined model class when loaded with \texttt{Rombus}. This is necessary to provide, for example, the priors and synthetic interferometers used for evaluating the frequency dependent antenna factors in the detector response. To read in the user-defined model, we pass the model file and keyword arguments \texttt{*kwargs} to
\begin{verbatim}
    model = RombusModel.load(model.py, *kwargs)
\end{verbatim}
The model can now be passed into the model-building elements of \texttt{Rombus}. First, though, we must build the downsampled domain and the training set.

We define the \texttt{utils.down\_sample} method to construct the downsampled domain based on the algorithm described in Sec.~\ref{sec:construc_training_set}. The method takes the {}\texttt{.ini} file used for \texttt{Bilby} inference as an argument and reads the duration, frequency range, minimum component masses and spins to resample the frequency domain. It also takes an optional argument \texttt{df\_max} which defines the maximum frequency resolution for down sampling. This is important to prevent errors accruing from the time-to-merger approximation breaking down at high frequencies. We downsample the model domain by setting 
\begin{verbatim}
    model.domain = utils.down_sample(filename.ini, df_max) 
\end{verbatim}

After setting the frequency domain, the training set is constructed. We instantiate the training set samples using 
\begin{verbatim}
    samples = Samples(model=model, n_random=1)
\end{verbatim}
Here, \texttt{Samples} is the base \texttt{Rombus} class to hold parameter samples. We populate the samples with a single sample as specified by \texttt{n\_random=1}. We define the methods \texttt{utils.boundary\_samples}, \texttt{utils.grid\_samples}, and \texttt{utils.random\_samples} to populate the training set based on the methods outlined in Sec.~\ref{sec:construc_training_set}. Only \texttt{model} and \texttt{n\_samples}, an integer for the number of samples, needs to be passed to the sampling methods. 

The first method provides all combinations of the minimum and maximum prior boundaries to include in the training set. The second and third method populate the training set with \texttt{n\_samples} samples taken from a uniform grid or randomly drawn from parameter space. The only parameter sampled from a non-uniform grid is the chirp mass $\mathcal{M}_c$, which is sampled uniformly in $\mathcal{M}_c^{-5/3}$ as this term describes the inspiral to leading order. This non-uniform sampling is also implemented in the \texttt{utils.grid\_samples} method. 

After obtaining the grid, random, and boundary samples, they are passed to the \texttt{samples} object by executing, for example,
\begin{verbatim}
    samples.extend(boundary_samples)
\end{verbatim}
To enable efficient parallelisation, we rewrote the \texttt{extend} method to scatter the samples across all CPUs used to build the training set and keep track of how many samples are on each CPU. To obtain an equal number of samples on each CPU requires adding at most one extra sample to each CPU, which is computationally negligible. 

Finally, the reduced order model is constructed by evaluating
\begin{verbatim}
    rom = ReducedOrderModel(model, samples, psd=psd).build(do_step=None, tol=tol)
\end{verbatim}
Here, \texttt{model} is the \texttt{RombusModel} instance that describes our full order model, \texttt{samples} is a \texttt{Samples} instance that we previously populated with parameter samples and is used to build the training set, and \texttt{psd} is the set of weights passed to the noise-weighted inner product. In the case of gravitational-wave astronomy, we set the weights to the power spectral density of the detector network. The \texttt{ReducedOrderModel} class is instantiated with these arguments and the \texttt{build} method is called. Setting \texttt{do\_step=None} tells \texttt{Rombus} to construct the reduced basis and the empirical interpolant. However, to construct one or the other simply set \texttt{do\_step="RB"} or \texttt{do\_step="EI"}, respectively. 

The keyword argument \texttt{tol} is the user-defined tolerance for the maximum reduced basis approximation error. Throughout this work, we set the tolerance to \texttt{tol=1e-16}. This tolerance is comparable to the numerical precision of a double precision floating point number, used to represent the real and imaginary parts of the complex numbers used in this work. Depending on the problem at hand, such extreme levels of precision is not necessary.  

In order to build almost any accurate reduced order model for gravitational-wave science, the reduced order modelling algorithm must be parallelised. A major contribution we have made to \texttt{Rombus} is parallelising the reduced order modelling process, allowing for the efficient construction of ROMs on high performance computing clusters. Typically, we parallelise the ROM construction over hundreds of CPUs.

Calling \texttt{build(do\_step=None, tol=tol)} creates the training set out of the samples provided in \texttt{samples}, use the training set to construct the reduced basis, and then use the reduced basis to construct the empirical interpolant. These three data structures are contained in the \texttt{rom} variable. So are the greedily selected parameters \texttt{rom.reduced\_basis.greedypoints}. These are the parameters used to obtain the waveforms in the reduced basis.

As described in Sec.~\ref{sec:construc_training_set}, these greedily selected parameters are integral to the up-sampling procedure of the reduced order model. Both the up-sampled linear and quadratic reduced order models are constructed over a training set consisting of \textit{only} the greedily selected parameter samples. Even if adaptive sampling is not used, the quadratic model is still only constructed over these samples. This technique allows the user to dodge the extreme computational expense that comes with making a large training set over a uniform frequency domain. 

In some cases, it is computationally impossible to make training sets large enough despite the adaptive sampling algorithm. In this case, we can increase the effective training set size --- and therefore ROM accuracy --- using the functionality 
\begin{verbatim}
        rom = rom.refine(N, tol=tol, iterate=False)
\end{verbatim}
The \texttt{rom.refine} method tells \texttt{Rombus} to create a new training set of \texttt{N} random parameter samples and to rerun the reduced basis algorithm on the existing basis but using this new training set. This has the effect of increasing the effective training set size by \texttt{N} samples, as described in Chapter~\ref{sec:construc_training_set}.

The keyword argument \texttt{iterate} dictates whether this refinement procedure proceeds until a completely new training set of \texttt{N} waveforms has no elements with approximation errors greater than \texttt{tol}. Setting \texttt{iterate=True} can cause the technique to iterate forever if using a tolerance comparable to machine precision due to numerical errors in the projection coefficients. Therefore, as we frequently use tolerances at this scale, we set \texttt{iterate=False} and loop over the refinement procedure \texttt{n\_refine=3} times. 

The components of the reduced order model that are actually used during inference are the interpolation matrix $B(f)$ and frequency nodes $\{F_i\}_{i=1}^{N}$ described in Sec.~\ref{sec:building_rom}. These quantities can be saved to an output directory for use in inference. They can be easily accessed as 
\begin{verbatim}
    rom.empirical_interpolant.B_matrix
    rom.empirical_interpolant.nodes
\end{verbatim}
To use a reduced order model for inference using \texttt{Bilby}, simply add the following lines to an \texttt{.ini} file: 
\begin{verbatim}
    likelihood-type = ROQGravitationalWaveTransient
    roq_folder = ROQ_data
\end{verbatim}
Here, \texttt{ROQGravitationalWaveTransient} is the standard reduced order quadrature likelihood implemented in \texttt{Bilby}. 

The reduced order quadrature likelihood used in this work (Eq.~\ref{eq:roq}) has been implemented in a forked version of \texttt{Bilby}, and is accessed by instead specifying
\begin{verbatim}
    likelihood-type = TimeDependentROQGravitationalWaveTransient
\end{verbatim}
The key difference in our implementation is that it allows the inclusion of the effects due to Earth's rotation and the free spectral range (FSR), which is important to model for next-generation detectors.
The standard \texttt{ROQGravitationalWaveTransient} likelihood implemented in \texttt{Bilby} does not include the effects due to Earths rotation and the FSR.

\section{Effects due to Earth's  Rotation and the Free Spectral Range}\label{sec:jacob}

Current ground-based gravitational wave detectors have arm lengths (L), which are much shorter than the wavelengths of the gravitational waves they are sensitive to ($\theta_{\rm GW}$). Equivalently, the characteristic frequency for light's round-trip along the arm, known as the free spectral range, $f_{\rm fsr} = c/{2L} \approx 37 \rm{kHz} >> f_{\rm GW}$. In this limit (the ``long wavelength approximation"), $f_{\rm GW}$ is approximately stationary along the light travel timescale, and the response of the detector to the signal is independent of $f_{\rm GW}$.

With $L = 40 \rm \ km$, CE will have an $f_{\rm fsr} \approx 3.7 \rm km$, which is much closer to the gravitational wave frequencies it will be sensitive to. Relaxing the long-wavelength approximation, we must now account for the fact that the change in light travel time along a detector arm has additional dependence on how many cycles of the gravitational wave the light interacts with (previously it was assumed that the light was affected by the same phase of the gravitational wave signal, if present, for the entire duration of its trip down the arms). Accounting for this therefore requires making the detector's response function explicitly depend on $f_{\rm GW}$ and the projection of the direction to the source onto the arm (see, e.g., \cite{Rakhmanov:2008is, Essick_2017}):

\begin{equation}
    D(f, n_a) = \frac{f_{\rm fsr}}{4 \pi i f} \big( \frac{1 - e^{\frac{-\pi i f (1-n_a)}{f_{\rm fsr}}}}{1-n_a} - e^{\frac{-2\pi i f}{f_{\rm fsr}}} \frac{1 - e^{\frac{\pi i (1+n_a) f}{f_{\rm fsr}}}}{1+n_a}\big),
\end{equation}
where $n_a \equiv \vec{n} \cdot \hat{e}_a$ is the line-of-sight vector to the source $\vec{n}$ projected onto the unit vector $\hat{e}_a$ which points along arm $a$. 

This form of $D$ is used in Eq.~\ref{eq:fpc} to compute $F^{+,\times}$.
In the long-wavelength approximation used in typical analyses, $D$ is a constant independent of $f$ and $\vec{n}$.

The duration of signals seen in current ground-based detectors is much shorter than the timescale on which $\vec{n}$ changes due to the rotation of Earth, making it safe to fix $\vec{n}$ when projecting a simulated signal into the detector. With 3G detectors seeing signals for longer durations, these two timescales become comparable. We must therefore account for how the variation of $\vec{n}$ while the signal is visible affects the sensitivity of the detector to the signal. 

While the equatorial/celestial coordinates (i.e., right ascension and declination) of $\vec{n}$ are fixed, Eq.~\ref{eq:fpc} must be evaluated in the frame of the detector, a coordinate system in which $\vec{n}$ now explicitly depends on the current angular position between the detector and the source, making Eq.~\ref{eq:fpc} depend on the current time. For long duration signals, this change in $\vec{n}$ means $D$ and $\epsilon$ also change while the signal is in band. Because Eq.~\ref{eq:fpc} is being evaluated on a grid of frequencies and $\vec{n}$ depends on the GPS time at the detector, we must calculate the time in the detector for each frequency point on the grid. In practice, we use a 1.5PN approximation for the time to merger to calculate $t(f) - t_c$ for each point in the frequency grid $f$ (e.g., \cite{Cutler_1994}).

In \texttt{bilby}, the signal at the detector location is calculated by first calculating $h(f)$ in the geocenter frame and then applying a time/phase shift to project the signal from the geocenter to the detector. This projection depends on the angle between $\vec{n}$ and the vector connecting the geocenter to the detector location. As this is a time-dependent projection for long duration signals, we correct for this Doppler effect in the phase by performing a time-dependent phase shift in the frequency domain:

\begin{equation}
    h(f) = h_{\rm geocent}(f) e^{-2 \pi i f \delta t(f)},
\end{equation}

where $\delta t (f)$ is the difference in time between when the part of the signal with frequency $f$ arrives in the geocenter and when it arrives at the detector. Whereas for analyses with current ground-based detectors, $\delta t$ is a constant independent of $f$, $\delta t$ is now an array with components determined by the time-to-merger approximation mentioned above.

\section{Plots and Tables}\label{app:plots}
\begin{figure}[h]
  \centering
  \includegraphics[scale=0.8]{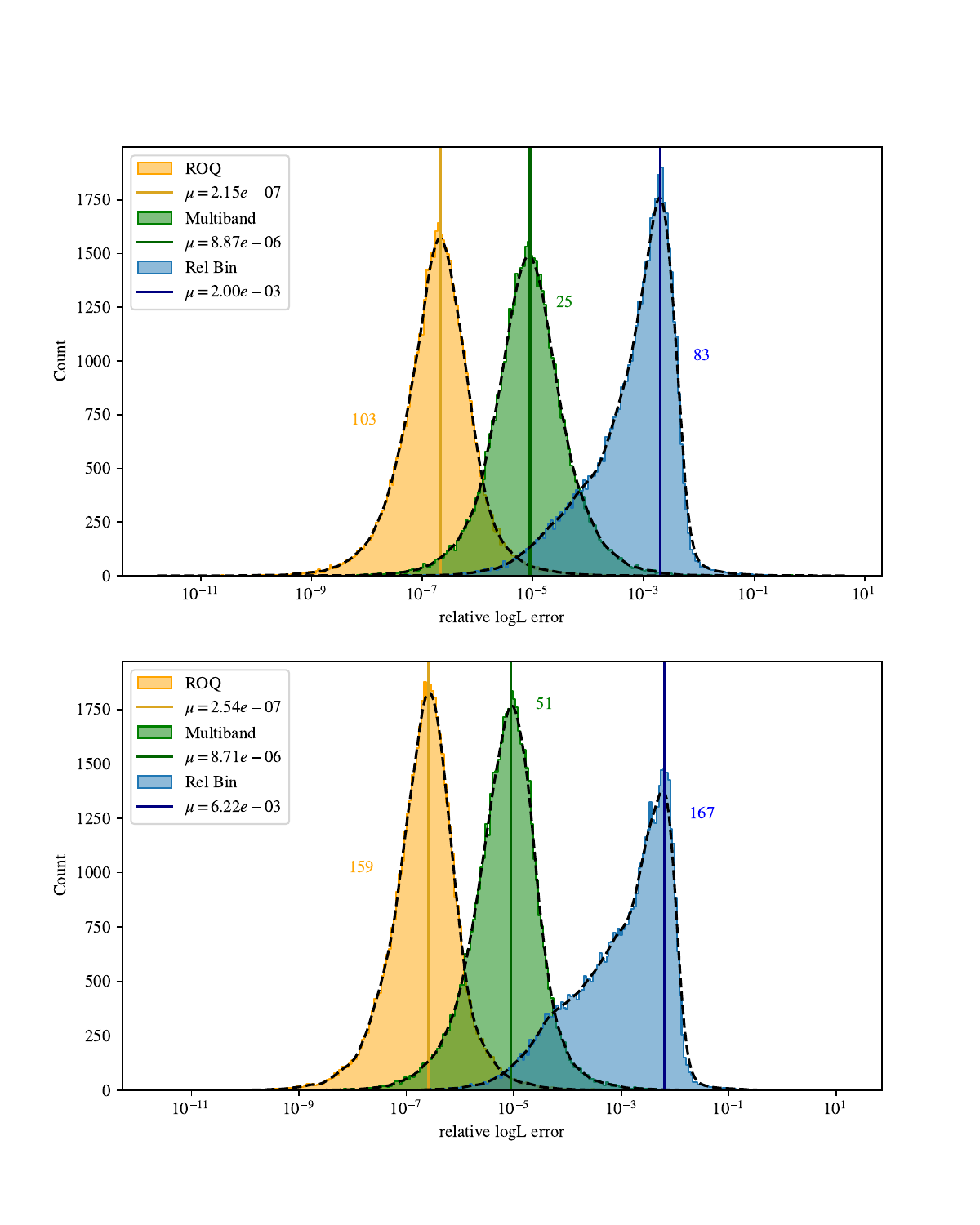}
  \caption{Relative log likelihood ratio errors for a fixed range in chirp mass $\Delta\mathcal{M}_c$ for a linear reduced order quadrature rule, multibanding, and relative binning. Both relative binning and multibanding ignore the effects due to Earth's rotation and the free spectral range, whereas the reduced order quadrature method includes them. The top panel compares the likelihood acceleration techniques over $50,000$ waveforms from $f_{\min}=\unit[30]{Hz}$  and chirp mass range $\Delta\mathcal{M}_c=6\times10^{-4}$. The bottom panel shows the same comparison with waveforms evaluated from $f_{\min}=\unit[24]{Hz}$ with chirp mass range $\Delta\mathcal{M}_c=9\times10^{-4}$. The peak of the error distributions are indicated in the figure legend, with the associated likelihood speedups listed next to each curve. For relative binning we set the tunable accuracy parameters $\chi, \epsilon$ to $\chi=10, \epsilon=0.1$, and for multi-banding we set the accuracy parameter $L$ to $L=5$. The accuracy of multi-banding and relative binning can be increased by modifying these accuracy parameters, but this has the effect of further slowing inference. }
  \label{fig:rb_roq_rel_ll_error}
\end{figure}

\begin{figure}
    \centering
    \includegraphics[width=\linewidth]{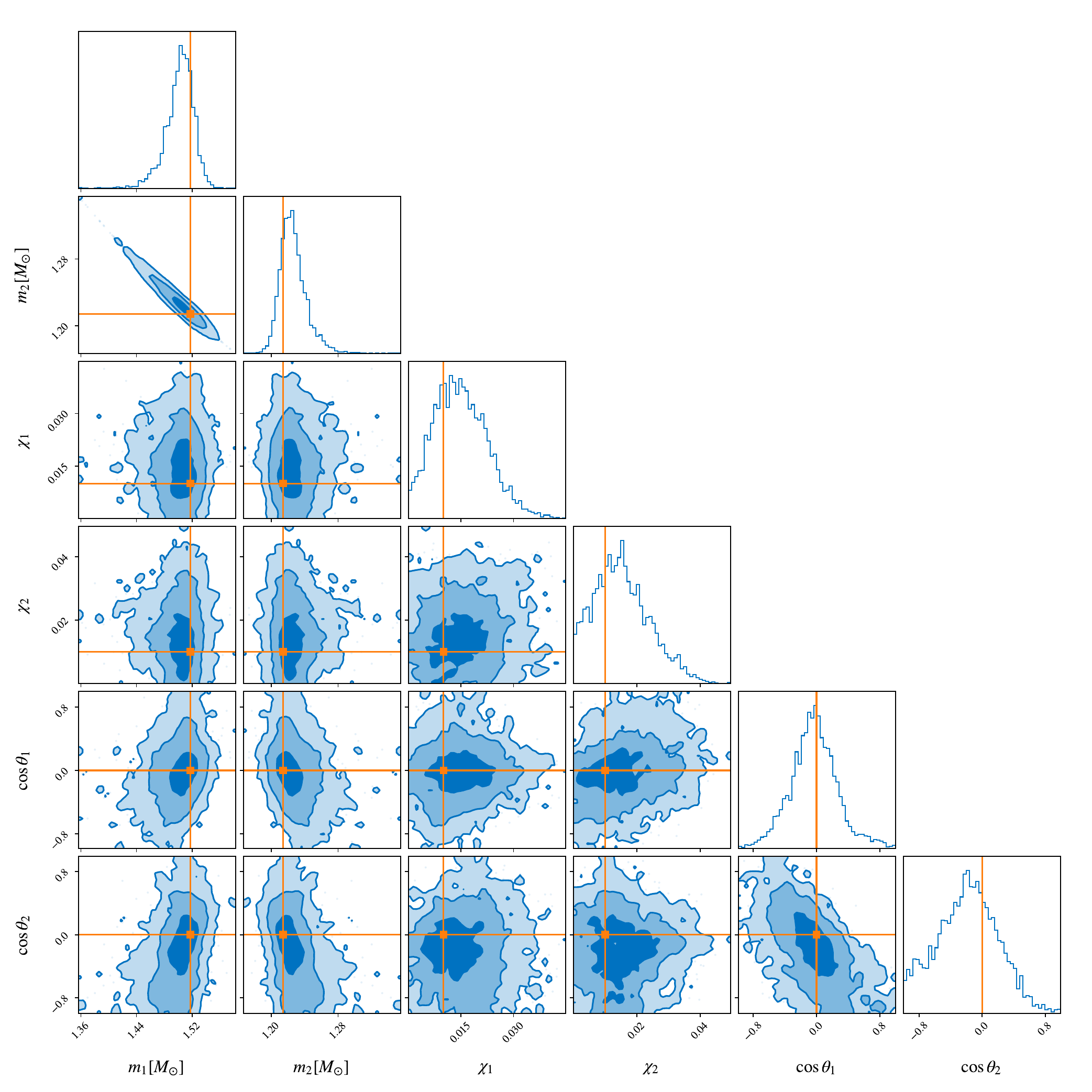}
    \caption{One- and two- dimensional posterior distributions for the component masses $m_{1,2} [M_{\odot}]$, spin magnitudes $\chi_{1,2}$, and cosine of the spin tilts $\cos\theta_{1,2}$ of a GW170817-like binary neutron star signal injected into a single Cosmic Explorer with SNR=792 and $f_\text{min}=\unit[24]{Hz}$.
    The true values are indicated in orange.
    }
    \label{fig:26hz_masses_spins}
\end{figure}

\begin{figure}
    \centering
    \includegraphics[width=0.5\linewidth]{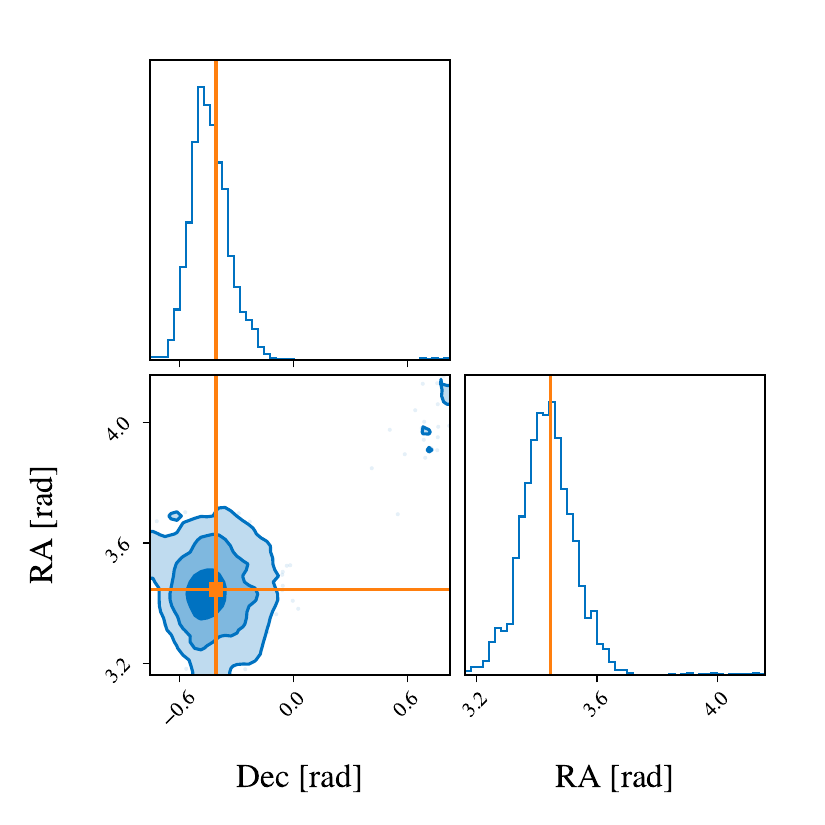}
    \caption{One- and two- dimensional posterior distributions for the right ascension $\text{RA} [rad]$ and declination $\text{Dec} [rad]$ of a GW170817-like binary neutron star signal injected into a single Cosmic Explorer with SNR=792 and $f_\text{min}=\unit[24]{Hz}$.
    The true values are indicated in orange.
    }
    \label{fig:26hz_source_location}
\end{figure}

\begin{figure}
    \centering
    \includegraphics[width=0.5\linewidth]{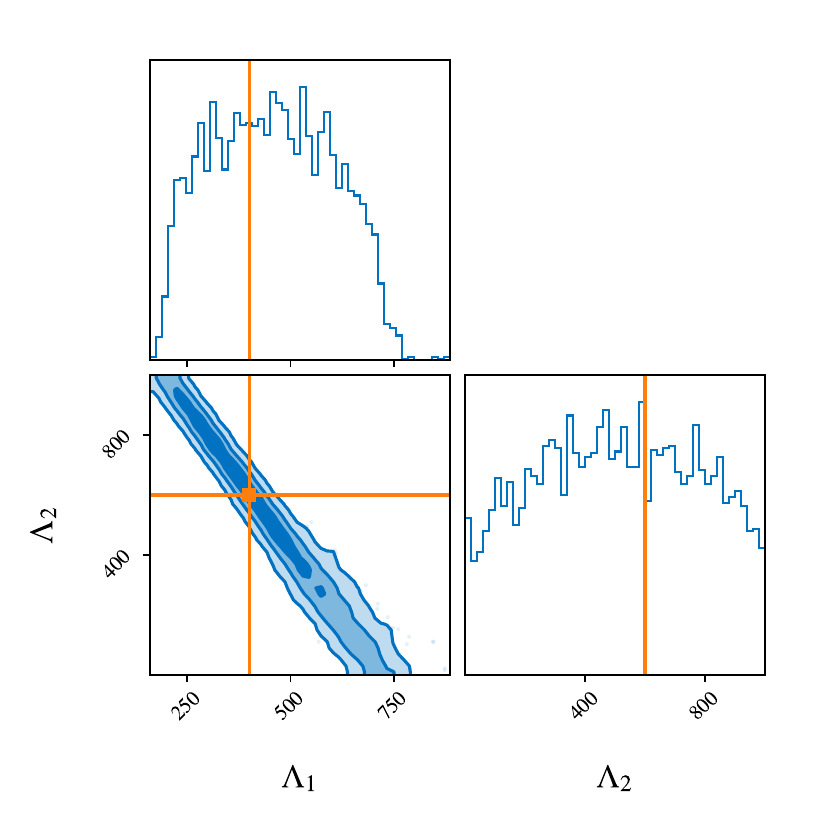}
    \caption{One- and two- dimensional posterior distributions for the tidal deformabilities $\Lambda_{1,2}$ of a GW170817-like binary neutron star signal injected into a single Cosmic Explorer with SNR=792 and $f_\text{min}=\unit[24]{Hz}$.
    The true values are indicated in orange.
    }
    \label{fig:26hz_tides_time}
\end{figure}

\begin{figure}[h]
    \centering
    \includegraphics[width=0.7\linewidth]{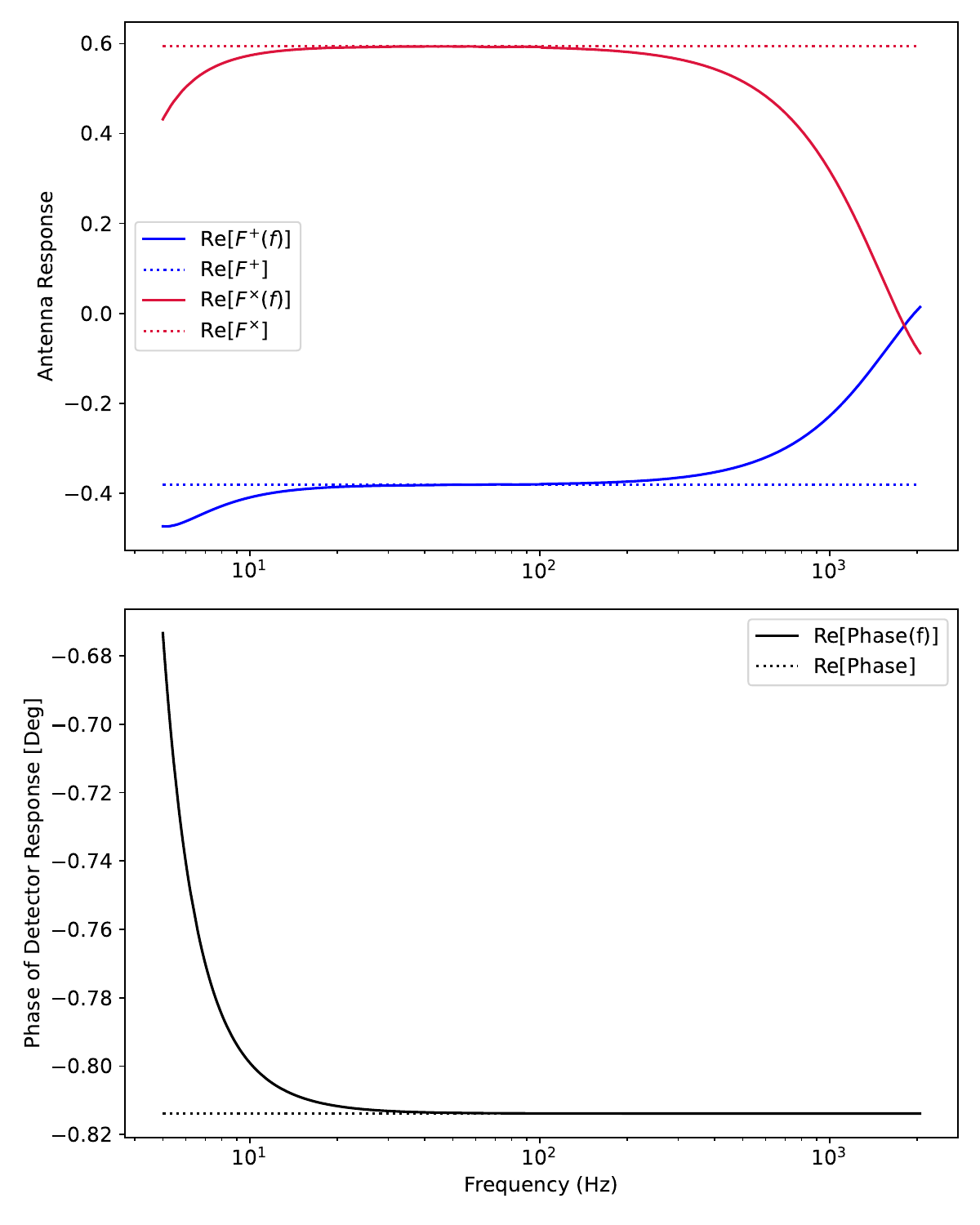}
    \caption{\textit{Top:} The real components of the plus- and cross-polarised antenna response factors in Eq.~\ref{eq:fpc}. \textit{Bottom:} The real component of the phase of the detector response as defined in Eq.~\ref{eq:phase}. All quantities are evaluated using Cosmic Explorer, a fixed source location at $(\text{dec, RA})=(0,0)$ and polarisation angle $\psi=0$, and illustrated as frequency-dependent (solid line) and frequency-independent (dashed line).
    }
    \label{fig:fp_phase_vs_constant}
\end{figure}

\begin{figure}[h]
    \centering
    \includegraphics[width=\linewidth]{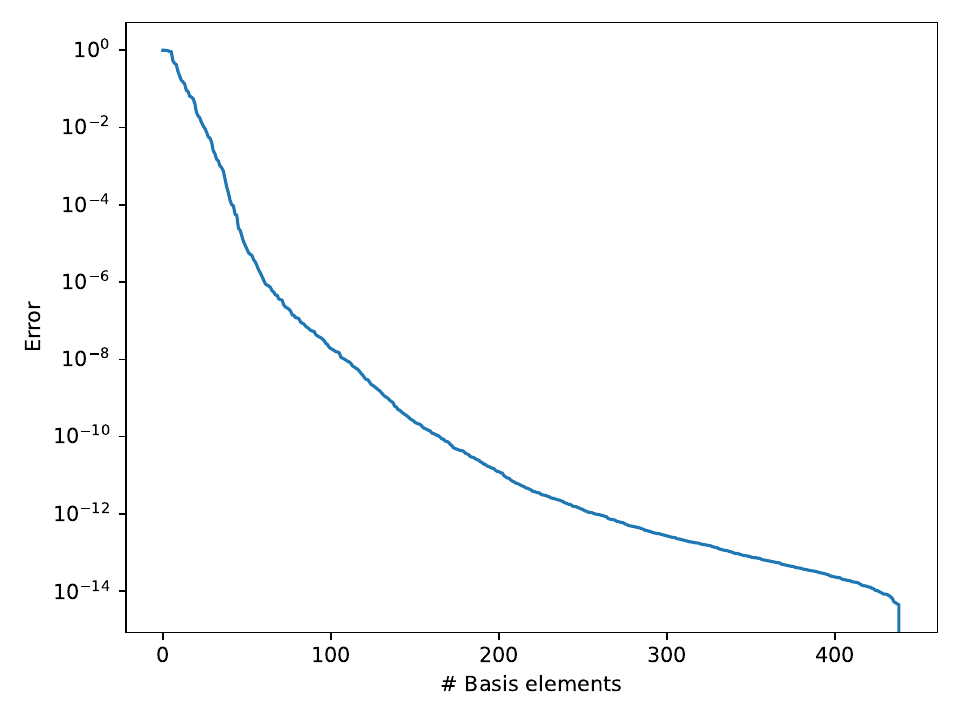}
    \caption[Reduced basis approximation error]{
    The maximum reduced basis approximation error onto a training set of 2 million binary neutron star merger signals as a function of the number of basis elements. The signals are constructed  with a minimum frequency of $f_{\min}=\unit[15]{Hz}$. 
    The tolerance  for the maximum approximation error is $\epsilon=10^{-16}$. 
    At this error scale, numerical precision errors dominate and the mismatch between waveforms can become negative. 
    This is why we observe a sharp drop at the $460^\text{th}$ basis element: it is computationally infeasible to make a basis that is any more accurate.}
    \label{fig:rb_error}
\end{figure}

This appendix includes Figs.~\ref{fig:rb_roq_rel_ll_error}-\ref{fig:rb_error}, which provide supporting details to the main body. Fig.~\ref{fig:rb_roq_rel_ll_error} shows the relative error in the log likelihood ratio when using reduced order quadrature, multibanding, and relative binning likelihood acceleration methods. Importantly, the reduced order quadrature rule used in Fig.~\ref{fig:rb_roq_rel_ll_error} includes the full set of effects due to Earth's rotation and the free spectral range, whereas the relative binning and multibanding methods do not. Figs.~\ref{fig:26hz_masses_spins}-~\ref{fig:26hz_tides_time} show the posterior distributions from inference performed on a GW170817-like binary neutron star signal injected into a single Cosmic Explorer with $\text{SNR}=792$ and from a minimum frequency of $f_{\min}=\unit[24]{Hz}$. We use an ROQ rule that includes the effects due to Earth's rotation and the free spectral range. The signal was injected into a zero-noise realisation of Gaussian noise, and therefore we expect the posterior distributions to peak at the injected parameters. In Fig.~\ref{fig:fp_phase_vs_constant} we plot the real parts of the frequency-dependent and frequency-independent antenna response factors and phase of the detector response. In Fig.~\ref{fig:rb_error} we plot the maximum reduced basis approximation error as a function of the number of basis elements over a training set of 2 million binary neutron star waveforms starting from $f_{\min}=15$Hz. Finally, in Table~\ref{tab:rom_summary} we describe the size and accuracy of the reduced order models plotted in Fig.~\ref{fig:snr_curve}.

\begin{table}[h]
    \centering
    \caption{Summary of ROMs in Fig.~\ref{fig:snr_curve}. Note that the ROM from $f_{\min}=\unit[15]{Hz}$ is built using a relaxed error tolerance in order to keep the basis size under $\unit[100]{Gb}$, whereas the other ROMs are built using the same error tolerance of $\epsilon=10^{-16}$.}
    \begin{tabular}{c|c|c|c}
    \hline 
    \hline 
    $f_{\min}$ (Hz) & \# Linear Basis Elements & Max SNR & $\Delta\mathcal{M}_c$ $(M_{\odot})$ \\
    15 & 2029 & 608 & $2.2\times10^{-3}$ \\ 
    22 & 3388 & 1723 & $1\times10^{-3}$\\ 
    24 & 3260 & 1793 & $9\times10^{-4}$\\ 
    26 & 2666 & 1928 & $7.5\times10^{-4}$\\ 
    28 & 2520 & 1973 & $7\times10^{-4}$\\ 
    30 & 2493 & 2157 & $6\times10^{-4}$\\
    \hline 
    \hline
    \end{tabular}
    
    \label{tab:rom_summary}
\end{table}

\section{Analysis Techniques}\label{appendix:aprior}
This Appendix describes two useful results that support the main body. First, we derive an approximate relation between the maximum SNR of a signal that can be studied without introducing biases and the relative error in the log likelihood ratio between the reduced order quadrature and standard likelihoods. Then, we demonstrate that it is computationally inefficient to construct ROMs for individual components of the detector response as opposed to a single ROM for the entire detector response.

\subsection{Approximate A Priori Error Bounds}
We derive a relation between the relative log likelihood ratio error and the maximum signal to noise ratio at which we can obtain unbiased inference. From \cite{Lindbolm_2008}, for unbiased parameter estimation we require $|h-h^\text{interp}|^2\leq 1$. Furthermore, the log likelihood ratio is on the order of the squared signal-to-noise ratio $|h|^2 = \langle h,h\rangle = \rho^2$ \citep{Morisaki_2023}. Therefore, the absolute relative error in log likelihood ratio to leading order is given by
\begin{align}
    \left|\frac{\delta \ln \mathcal{L}}{\ln \mathcal{L}}\right| &\approx \frac{\left|\left|h\right|^2 - \left|h^\text{interp}\right|^2\right|}{\left|h\right|^2}. 
\end{align}
Here, $\delta \ln \mathcal{L} = \ln\mathcal{L} - \ln\mathcal{L}_{\text{ROQ}}$. From the triangle inequality $||a|-|b||\leq |a-b|$, we have that $|a|^2+|b|^2-2|a||b| \leq |a-b|^2$. We can write 
\begin{align}
    \left|\frac{\delta \ln \mathcal{L}}{\ln \mathcal{L}}\right| &= \frac{\left|
    \left|h\right|^2 + \left|h^\text{interp}\right|^2 - 2\left|h^\text{interp}\right|^2\right|}{\left|h\right|^2}. 
\end{align}

If we assume that $|h|\approx |h^\text{interp}|$ (but $|h|^2\not\approx|h^\text{interp}|^2$), then 
\begin{align}
    \left|\frac{\delta \ln \mathcal{L}}{\ln \mathcal{L}}\right| &\approx \frac{\left|\left|h\right|^2 + \left|h^\text{interp}\right|^2 - 2\left|h^\text{interp}\right|\left|h\right|\right|}{\left|h\right|^2} \\ 
    &\leq \frac{\left|h-h^\text{interp}\right|^2}{\left|h\right|^2} \\ 
    &\leq \frac{1}{\rho^2}.
\end{align}
Here, we have made use of the triangle inequality and Lindblom's requirement. Therefore, the optimal signal-to-noise ratio sets an approximate upper bound on the relative error in the log likelihood ratio required for unbiased inference.

\subsection{Choice of Reduced Order Quadrature} \label{appendix:roq_choice}
The greedy nature of the reduced basis algorithm implies that building separate bases for the components of the detector response will not yield greater compression than simply building a basis for the detector response itself. This is easily illustrated by considering separate models for the components of the phase-independent response, namely $h^{+/\times}$ and $F^{+/\times}$. Assuming these reduced order models approximate $F^{\times}$ and $h^{\times}$ well, then we have 
\begin{align}
    F^{+/\times}(f) &= \sum_a F^{+/\times}(F_a) A_a \\ 
    h^{+/\times}(f) &= \sum_b h^{+/\times}(F_b) B_b.
\end{align}
The log likelihood ratio is then
\begin{align}
    \ln \mathcal{L} &= -\langle F^+h^+, F^{\times} h^{\times}\rangle + \sum_{p=+/\times}\Big(\langle d, F^ph^p \rangle - \frac{1}{2}\langle F^ph^p, F^ph^p \rangle\Big).
\end{align}
If we consider the simplest term, $\langle d, F^+h^+ \rangle$, we see that
\begin{align}
    \langle d, F^+h^+ \rangle &= \sum_a\sum_b\sum_i \frac{d_i F^+(F_a)A_a(f_i) h^+(F_b)B_b(f_i)}{P_i} \\ 
    &= \sum_a\sum_b F^+(F_a)h^+(F_b) \psi_{ab}.
\end{align}
Here, $d_i$ is the gravitational wave data, $P_i$ is the power spectral density of the detector, and $\{f_i\}_i^N$ is the discrete frequency domain over which the inference takes place. Furthermore, $\psi_{ab}$ are the weights 
\begin{align}
    \psi_{ab} &= \sum_i \frac{d_i A_a(f_i)B_b(f_i)}{P_i}.
\end{align}
It is obvious that the same nested sum emerges for the remaining 4 sums in the log likelihood ratio. Therefore, the log likelihood ratio would scale as $5ab$. Given the greedy nature of the reduced basis algorithm, we can assume $ab \geq N_L$, where $N_L$ is the size of the linear reduced basis required to represent $F^{p}h^{p}\big|_{p=+/\times}$. We can justify this as follows.

Consider the basis elements required to represent $F^+$, given by $A=\{A_i\}_{i=1}^a$, and the basis elements required to represent $h^+$, given by $B=\{B_j\}_{j=1}^b$. If $F^+$ and $h^+$ share no information, then $F^+$ is not in the span of $B$. Then the basis required to represent $F^+h^+$, given by $C=\{C_k\}_{k=1}^c$ with cardinality $N_L$, is simply 
\begin{align}
    C=\{ \alpha\beta | \alpha \in A, \beta\in B\},
\end{align}
and therefore $N_L = ab$. However, if $F^+$ and $h^+$ share information, i.e. $F^+$ is in the span of $B$ for some parameters $\theta$ (as $F^+\equiv F^+(\theta)$), then 
\begin{align}
    C\subseteq\{ \alpha\beta | \alpha \in A, \beta\in B\},
\end{align}
and has cardinality $N_L \leq ab$. This is a result of the spans of $A$ and $B$ overlapping. 

The size of the quadratic basis is in general smaller than the linear basis as we no longer model the detector response phase. Therefore, we have $N_Q \leq N_L$ and we arrive at $5ab > 2N_L \geq N_L + N_Q$. So, constructing bases for components of the detector response is computationally inefficient compared to a single basis for the entire detector response.

\bibliography{references}

\begin{thebibliography}{}
\expandafter\ifx\csname natexlab\endcsname\relax\def\natexlab#1{#1}\fi
\providecommand{\url}[1]{\href{#1}{#1}}
\providecommand{\dodoi}[1]{doi:~\href{http://doi.org/#1}{\nolinkurl{#1}}}
\providecommand{\doeprint}[1]{\href{http://ascl.net/#1}{\nolinkurl{http://ascl.net/#1}}}
\providecommand{\doarXiv}[1]{\href{https://arxiv.org/abs/#1}{\nolinkurl{https://arxiv.org/abs/#1}}}

\bibitem[{Abbott {et~al.}(2020{\natexlab{a}})Abbott, Abbott, Abbott, Abraham, \& Acernese}]{gw190425}
Abbott, B.~P., Abbott, R., Abbott, T.~D., Abraham, S., \& Acernese, F. 2020{\natexlab{a}}, The Astrophysical Journal Letters, 892, L3

\bibitem[{Abbott {et~al.}(2017)Abbott, Abbott, Abbott, Acernese, \& Ackley}]{GW170817}
Abbott, B.~P., Abbott, R., Abbott, T.~D., Acernese, F., \& Ackley, K. 2017, Phys. Rev. Lett., 119, 161101

\bibitem[{Abbott {et~al.}(2018)}]{GW170817_stoch}
Abbott, B.~P., {et~al.} 2018, Phys. Rev. Lett., 120, 091101

\bibitem[{Abbott {et~al.}(2019)Abbott, Abbott, Abbott, Abraham, Acernese, Ackley, Adams, Adhikari, Adya, \& Affeldt}]{gwtc1}
Abbott, B.~P., Abbott, R., Abbott, T.~D., {et~al.} 2019, Phys. Rev. X, 9, 031040, \dodoi{10.1103/PhysRevX.9.031040}

\bibitem[{Abbott {et~al.}(2020{\natexlab{b}})}]{ligo_noise}
Abbott, B.~P., {et~al.} 2020{\natexlab{b}}, Class. Quantum Grav., 37, 055002

\bibitem[{Abbott {et~al.}(2023)Abbott, Abbott, Acernese, Ackley, Adams, Adhikari, Adhikari, \& Adya}]{gwtc3}
Abbott, R., Abbott, T.~D., Acernese, F., {et~al.} 2023, Phys. Rev. X, 13, 041039, \dodoi{10.1103/PhysRevX.13.041039}

\bibitem[{Abbott {et~al.}(2021)Abbott, Abbott, Abraham, Acernese, Ackley, Adams, Adams, Adhikari, \& Adya}]{gwtc2}
Abbott, R., Abbott, T.~D., Abraham, S., {et~al.} 2021, Phys. Rev. X, 11, 021053, \dodoi{10.1103/PhysRevX.11.021053}

\bibitem[{Ashton \& Talbot(2021)}]{bilbymcmc}
Ashton, G., \& Talbot, C. 2021, Monthly Notices of the Royal Astronomical Society, 507, 2037, \dodoi{10.1093/mnras/stab2236}

\bibitem[{Ashton {et~al.}(2019)}]{bilby}
Ashton, G., {et~al.} 2019, Astrophys. J. Supp., 241, 27

\bibitem[{Baral {et~al.}(2023)Baral, Morisaki, Hernandez, \& Creighton}]{baral}
Baral, P., Morisaki, S., Hernandez, I. M.~n., \& Creighton, J. 2023, Phys. Rev. D, 108, 043010, \dodoi{10.1103/PhysRevD.108.043010}

\bibitem[{Barrault {et~al.}(2004)Barrault, Maday, Nguyen, \& Patera}]{Barrault_2004}
Barrault, M., Maday, Y., Nguyen, N.~C., \& Patera, A.~T. 2004, Reports. Mathematical, 339, 667, \dodoi{10.1016/j.crma.2004.08.006}

\bibitem[{Biscoveanu {et~al.}(2020{\natexlab{a}})Biscoveanu, Haster, Vitale, \& Davies}]{Biscoveanu_2020}
Biscoveanu, S., Haster, C.-J., Vitale, S., \& Davies, J. 2020{\natexlab{a}}, Phys. Rev. D, 102

\bibitem[{Biscoveanu {et~al.}(2020{\natexlab{b}})Biscoveanu, Talbot, Thrane, \& Smith}]{cosmo}
Biscoveanu, S., Talbot, C., Thrane, E., \& Smith, R. 2020{\natexlab{b}}, Phys. Rev. Lett., 125, 241101

\bibitem[{Canizares {et~al.}(2013)Canizares, Field, Gair, \& Tiglio}]{Canizares_2013}
Canizares, P., Field, S.~E., Gair, J.~R., \& Tiglio, M. 2013, Phys. Rev. D, 87, 124005

\bibitem[{Chatziioannou {et~al.}(2021)Chatziioannou, Cornish, Wijngaarden, \& Littenberg}]{glitch_mitigations2}
Chatziioannou, K., Cornish, N., Wijngaarden, M., \& Littenberg, T.~B. 2021, Phys. Rev. D, 103, 044013

\bibitem[{Chatziioannou {et~al.}(2024)Chatziioannou, Dent, Fishbach, Ohme, Pürrer, Raymond, \& Veitch}]{chatziioannou2024compactbinarycoalescencesgravitationalwave}
Chatziioannou, K., Dent, T., Fishbach, M., {et~al.} 2024, Compact binary coalescences: gravitational-wave astronomy with ground-based detectors.
\newblock \doarXiv{2409.02037}

\bibitem[{Chen {et~al.}(2021)Chen, Cowperthwaite, Metzger, \& Berger}]{Chen_2021}
Chen, H.-Y., Cowperthwaite, P.~S., Metzger, B.~D., \& Berger, E. 2021, The Astrophysical Journal Letters, 908, L4

\bibitem[{Cornish(2010)}]{Cornish:2010kf}
Cornish, N.~J. 2010.
\newblock \doarXiv{1007.4820}

\bibitem[{Cutler \& Flanagan(1994)}]{Cutler_1994}
Cutler, C., \& Flanagan, E.~E. 1994, Phys. Rev. D, 49, 2658

\bibitem[{Dax {et~al.}(2021)Dax, Green, Gair, Macke, Buonanno, \& Sch{\"o}lkopf}]{dingo}
Dax, M., Green, S.~R., Gair, J., {et~al.} 2021, Phys. Rev. Lett., 127, 241103

\bibitem[{Dax {et~al.}(2024)Dax, Green, Gair, Gupte, Pürrer, Raymond, Wildberger, Macke, Buonanno, \& Schölkopf}]{dax2024realtimegravitationalwaveinferencebinary}
---. 2024, Real-time gravitational-wave inference for binary neutron stars using machine learning.
\newblock \doarXiv{2407.09602}

\bibitem[{DeVore {et~al.}(2013)DeVore, Petrova, \& Wojtaszczyk}]{devore}
DeVore, R., Petrova, G., \& Wojtaszczyk, P. 2013, Constructive Approximation, \dodoi{https://doi.org/10.1007/s00365-013-9186-2}

\bibitem[{Dietrich {et~al.}(2019)Dietrich, Samajdar, Khan, Johnson-McDaniel, Dudi, \& Tichy}]{Dietrich:2019kaq}
Dietrich, T., Samajdar, A., Khan, S., {et~al.} 2019, Phys. Rev. D, 100, 044003, \dodoi{10.1103/PhysRevD.100.044003}

\bibitem[{Essick(2022)}]{essick_2022}
Essick, R. 2022, Phys. Rev. D, 105, 082002

\bibitem[{Essick {et~al.}(2017)Essick, Vitale, \& Evans}]{Essick_2017}
Essick, R., Vitale, S., \& Evans, M. 2017, Phys. Rev. D, 96, 084004

\bibitem[{Field {et~al.}(2014)Field, Galley, Hesthaven, Kaye, \& Tiglio}]{Field}
Field, S.~E., Galley, C.~R., Hesthaven, J.~S., Kaye, J., \& Tiglio, M. 2014, Phys. Rev. X, 4, 031006

\bibitem[{Gardner {et~al.}(2023)}]{Gardner}
Gardner, J.~W., {et~al.} 2023, Phys. Rev. D, 108, 123026

\bibitem[{Greene {et~al.}(2020)Greene, Strader, \& Ho}]{Greene}
Greene, J.~E., Strader, J., \& Ho, L.~C. 2020, Ann. Rev. Astron. Astrophys., 58, 259

\bibitem[{Guttman {et~al.}(2025)Guttman, Lasky, \& Thrane}]{tPowerBilby}
Guttman, N., Lasky, P.~D., \& Thrane, E. 2025

\bibitem[{Hall {et~al.}(2021)Hall, Kuns, Smith, Bai, \& Wipf}]{Hall_2021}
Hall, E.~D., Kuns, K., Smith, J.~R., Bai, Y., \& Wipf, C. 2021, Phys. Rev. D, 103, 122004

\bibitem[{Hu {et~al.}(2024)Hu, Irwin, Sun, Messenger, Suleiman, Heng, \& Veitch}]{hu2024decodinglongdurationgravitationalwaves}
Hu, Q., Irwin, J., Sun, Q., {et~al.} 2024, Decoding Long-duration Gravitational Waves from Binary Neutron Stars with Machine Learning: Parameter Estimation and Equations of State.
\newblock \doarXiv{2412.03454}

\bibitem[{Hu \& Veitch(2024)}]{Hu}
Hu, Q., \& Veitch, J. 2024

\bibitem[{Iacovelli {et~al.}(2022)Iacovelli, Mancarella, Foffa, \& Maggiore}]{Iacovelli_2022}
Iacovelli, F., Mancarella, M., Foffa, S., \& Maggiore, M. 2022, The Astrophysical Journal, 941, 208

\bibitem[{Johnson {et~al.}(2024)Johnson, Chatziioannou, \& Farr}]{Johnson}
Johnson, A.~D., Chatziioannou, K., \& Farr, W.~M. 2024, Phys. Rev. D, 109, 084015

\bibitem[{Krishna {et~al.}(2023)Krishna, Vijaykumar, Ganguly, Talbot, Biscoveanu, George, Williams, \& Zimmerman}]{krishna2023acceleratedparameterestimationbilby}
Krishna, K., Vijaykumar, A., Ganguly, A., {et~al.} 2023, Accelerated parameter estimation in Bilby with relative binning.
\newblock \doarXiv{2312.06009}

\bibitem[{Leslie {et~al.}(2021)Leslie, Dai, \& Pratten}]{modebymode_relativebinning}
Leslie, N., Dai, L., \& Pratten, G. 2021, Phys. Rev. D, 104, 123030

\bibitem[{Lindblom {et~al.}(2008)Lindblom, Owen, \& Brown}]{Lindbolm_2008}
Lindblom, L., Owen, B.~J., \& Brown, D.~A. 2008, Phys. Rev. D, 78, 124020

\bibitem[{Martini~A.(2024)}]{Maximum_entropy}
Martini~A., Schmidt~S., A. G. e.~a. 2024, Eur. Phys. J. C, 84, 1023

\bibitem[{Morisaki(2021)}]{multibanding}
Morisaki, S. 2021, Phys. Rev. D, 104, 044062

\bibitem[{Morisaki {et~al.}(2023)Morisaki, Smith, Tsukada, Sachdev, Stevenson, Talbot, \& Zimmerman}]{Morisaki_2023}
Morisaki, S., Smith, R., Tsukada, L., {et~al.} 2023, Phys. Rev. D, 108, 123040, \dodoi{10.1103/PhysRevD.108.123040}

\bibitem[{Owen {et~al.}(2023)Owen, Haster, Perkins, Cornish, \& Yunes}]{Owen}
Owen, C.~B., Haster, C.-J., Perkins, S., Cornish, N.~J., \& Yunes, N. 2023, Phys. Rev. D, 108, 044018

\bibitem[{Plunkett {et~al.}(2022)Plunkett, Hourihane, \& Chatziioannou}]{PSD_effect_on_PE_1}
Plunkett, C., Hourihane, S., \& Chatziioannou, K. 2022, Phys. Rev. D, 106, 104021

\bibitem[{Punturo {et~al.}(2010)Punturo, Abernathy, Acernese, Allen, \& Andersson}]{Punturo_2010}
Punturo, M., Abernathy, M., Acernese, F., Allen, B., \& Andersson, N. 2010, Classical and Quantum Gravity, 27, 194002

\bibitem[{P\"urrer \& Haster(2020)}]{PhysRevResearch.2.023151}
P\"urrer, M., \& Haster, C.-J. 2020, Phys. Rev. Res., 2, 023151, \dodoi{10.1103/PhysRevResearch.2.023151}

\bibitem[{Rakhmanov {et~al.}(2008)Rakhmanov, Romano, \& Whelan}]{Rakhmanov:2008is}
Rakhmanov, M., Romano, J.~D., \& Whelan, J.~T. 2008, Class. Quant. Grav., 25, 184017, \dodoi{10.1088/0264-9381/25/18/184017}

\bibitem[{{Reitze} {et~al.}(2019){Reitze}, {Adhikari}, {Ballmer}, {Barish}, \& {Barsotti}}]{CE}
{Reitze}, D., {Adhikari}, R.~X., {Ballmer}, S., {Barish}, B., \& {Barsotti}, L. 2019, in Bulletin of the American Astronomical Society, Vol.~51, 35

\bibitem[{Romero-Shaw {et~al.}(2022)Romero-Shaw, Thrane, \& Lasky}]{wmf}
Romero-Shaw, I.~M., Thrane, E., \& Lasky, P.~D. 2022, Pub. Astron. Soc. Aust., 39, E025

\bibitem[{Romero-Shaw {et~al.}(2020)Romero-Shaw, Talbot, Biscoveanu, D’Emilio, Ashton, Berry, Coughlin, Galaudage, Hoy, Hübner, Phukon, Pitkin, Rizzo, Sarin, Smith, Stevenson, Vajpeyi, Arène, Athar, Banagiri, Bose, Carney, Chatziioannou, Clark, Colleoni, Cotesta, Edelman, Estellés, García-Quirós, Ghosh, Green, Haster, Husa, Keitel, Kim, Hernandez-Vivanco, Magaña~Hernandez, Karathanasis, Lasky, De~Lillo, Lower, Macleod, Mateu-Lucena, Miller, Millhouse, Morisaki, Oh, Ossokine, Payne, Powell, Pratten, Pürrer, Ramos-Buades, Raymond, Thrane, Veitch, Williams, Williams, \& Xiao}]{isobel_2021}
Romero-Shaw, I.~M., Talbot, C., Biscoveanu, S., {et~al.} 2020, Monthly Notices of the Royal Astronomical Society, 499, 3295, \dodoi{10.1093/mnras/staa2850}

\bibitem[{Rozza {et~al.}(2008)Rozza, Huynh, \& Patera}]{rb_pde}
Rozza, G., Huynh, D., \& Patera, A. 2008, Archives of Computational Methods in Engineering, \dodoi{https://doi.org/10.1007/s11831-008-9019-9}

\bibitem[{Santamar\'{\i}a {et~al.}(2010)Santamar\'{\i}a, Ohme, Ajith, Br\"ugmann, Dorband, Hannam, Husa, M\"osta, Pollney, Reisswig, Robinson, Seiler, \& Krishnan}]{Santamaria_2010}
Santamar\'{\i}a, L., Ohme, F., Ajith, P., {et~al.} 2010, Phys. Rev. D, 82, 064016, \dodoi{10.1103/PhysRevD.82.064016}

\bibitem[{Smith {et~al.}(2021)Smith, Borhanian, Sathyaprakash, Hernandez~Vivanco, \& Field}]{rory}
Smith, R., Borhanian, S., Sathyaprakash, B., Hernandez~Vivanco, F., \& Field, S.~E. 2021, Phys. Rev. Lett., 127, 081102

\bibitem[{Smith {et~al.}(2016)Smith, Field, Blackburn, Haster, P\"urrer, Raymond, \& Schmidt}]{Smith_2016}
Smith, R., Field, S.~E., Blackburn, K., {et~al.} 2016, Phys. Rev. D, 94, 044031

\bibitem[{Speagle(2020)}]{dynesty}
Speagle, J.~S. 2020, Monthly Notices of the Royal Astronomical Society, 493, 3132

\bibitem[{Talbot \& Thrane(2020)}]{student-t}
Talbot, C., \& Thrane, E. 2020, Phys. Rev. Res., 2, 043298

\bibitem[{Talbot {et~al.}(2021)Talbot, Thrane, Biscoveanu, \& Smith}]{windows}
Talbot, C., Thrane, E., Biscoveanu, S., \& Smith, R. 2021, Phys. Rev. Res., 3, 043049

\bibitem[{Thrane \& Talbot(2019)}]{thrane_talbot_2019}
Thrane, E., \& Talbot, C. 2019, Publications of the Astronomical Society of Australia, 36, e010, \dodoi{10.1017/pasa.2019.2}

\bibitem[{Zackay {et~al.}(2018)Zackay, Dai, \& Venumadhav}]{zackay2018relative}
Zackay, B., Dai, L., \& Venumadhav, T. 2018, Relative Binning and Fast Likelihood Evaluation for Gravitational Wave Parameter Estimation.
\newblock \doarXiv{1806.08792}

\end{thebibliography}

\end{document}